\documentclass[iop]{emulateapj}
\usepackage{aas_macros}

\usepackage{color}
\usepackage[dvipsnames,svgnames]{xcolor}

\usepackage{natbib}

\usepackage{amsmath}
\usepackage{xspace}
\usepackage{lineno}

\newcommand{\eg}{e.g.\xspace}

\mathchardef\mhyphen="2D

\newlength{\dhatheight}

\newcommand{\code}[1]{\texttt{#1}\xspace}

\newcommand{\unit}[1]{\ensuremath{\mathrm{\,#1}}\xspace}

\newcommand{\km}{\unit{km}}
\newcommand{\pc}{\unit{pc}}
\newcommand{\kpc}{\unit{kpc}}

\newcommand{\second}{\unit{s}}

\newcommand{\msun}{\unit{M_\odot}}

\newcommand{\lsun}{\unit{L_\odot}}

\newcommand{\Jfactor}{J-factor\xspace}
\newcommand{\Jfactors}{J-factors\xspace}

\providecommand\physrep{\ref@jnl{Phys.~Rep.}}%
\providecommand\apjs{\ref@jnl{ApJS}}%
\providecommand{\jcap}{\ref@jnl{JCAP}}%

\newcommand{\feh}         {\mbox{[Fe/H]}}
\newcommand{\kms}         {\ensuremath{\km\,\second^{-1}}\xspace}

\def\spose#1{\hbox to 0pt{#1\hss}}
\def\lta{\mathrel{\spose{\lower 3pt\hbox{$\mathchar"218$}}
     \raise 2.0pt\hbox{$\mathchar"13C$}}}
\def\gta{\mathrel{\spose{\lower 3pt\hbox{$\mathchar"218$}}
    \raise 2.0pt\hbox{$\mathchar"13E$}}}

\newcommand{\vbulk}{\ensuremath{75.6 \pm 1.3~\mbox{(stat.)} \pm 2.0~\mbox{(sys.)}}\xspace}
\newcommand{\vbulknoerr}{\ensuremath{75.6}\xspace}
\newcommand{\vgsr}{\ensuremath{-66.6}\xspace}
\newcommand{\vdisp}{\ensuremath{6.9^{+1.2}_{-0.9} }\xspace}
\newcommand{\mass}{\ensuremath{1.2^{+0.4}_{-0.3} \times 10^{7}}\xspace}
\newcommand{\masstolight}{\ensuremath{420^{+210}_{-140}}\xspace}
\newcommand{\fehmedian}{\ensuremath{-2.38 \pm 0.13}\xspace}
\newcommand{\fehdisp}{\ensuremath{0.47 ^{+0.12}_{-0.09}}\xspace}
\newcommand{\jsmall}{\ensuremath{16.5 \pm 0.8}\xspace}
\newcommand{\jlarge}{\ensuremath{16.6 \pm 0.9}\xspace}

\shorttitle{The Distant Milky Way Satellite Eridanus~II}
\shortauthors{Li et~al.}

\begin{document}

\title{Farthest Neighbor: The Distant Milky Way Satellite Eridanus~II\altaffilmark{*}}

\altaffiltext{*}{This paper includes data gathered with the 6.5 meter
  Magellan Telescopes located at Las Campanas Observatory, Chile.}


\def\andname{}

\author{
T.~S.~Li\altaffilmark{1,2},
J.~D.~Simon\altaffilmark{3},
A.~Drlica-Wagner\altaffilmark{1},
K.~Bechtol\altaffilmark{4,5,6},
M.~Y.~Wang\altaffilmark{2},
J.~Garc\'ia-Bellido\altaffilmark{7},
J.~Frieman\altaffilmark{1,8},
J.~L.~Marshall\altaffilmark{2},
D.~J.~James\altaffilmark{9,10},
L.~Strigari\altaffilmark{2},
A. B.~Pace\altaffilmark{2},
E.~Balbinot\altaffilmark{11},
Y.~Zhang\altaffilmark{1},
T. M. C.~Abbott\altaffilmark{10},
S.~Allam\altaffilmark{1},
A.~Benoit-L{\'e}vy\altaffilmark{12,13,14},
G.~M.~Bernstein\altaffilmark{15},
E.~Bertin\altaffilmark{12,14},
D.~Brooks\altaffilmark{13},
D.~L.~Burke\altaffilmark{16,17},
A. Carnero Rosell\altaffilmark{18,19},
M.~Carrasco~Kind\altaffilmark{20,21},
J.~Carretero\altaffilmark{22,23},
C.~E.~Cunha\altaffilmark{16},
C.~B.~D'Andrea\altaffilmark{24,25},
L.~N.~da Costa\altaffilmark{18,19},
D.~L.~DePoy\altaffilmark{2},
S.~Desai\altaffilmark{26},
H.~T.~Diehl\altaffilmark{1},
T.~F.~Eifler\altaffilmark{27},
B.~Flaugher\altaffilmark{1},
D.~A.~Goldstein\altaffilmark{28,29},
D.~Gruen\altaffilmark{16,17},
R.~A.~Gruendl\altaffilmark{20,21},
J.~Gschwend\altaffilmark{18,19},
G.~Gutierrez\altaffilmark{1},
E.~Krause\altaffilmark{16},
K.~Kuehn\altaffilmark{30},
H.~Lin\altaffilmark{1},
M.~A.~G.~Maia\altaffilmark{18,19},
M.~March\altaffilmark{15},
F.~Menanteau\altaffilmark{20,21},
R.~Miquel\altaffilmark{31,23},
A.~A.~Plazas\altaffilmark{27},
A.~K.~Romer\altaffilmark{32},
E.~Sanchez\altaffilmark{7},
B.~Santiago\altaffilmark{33,18},
M.~Schubnell\altaffilmark{34},
I.~Sevilla-Noarbe\altaffilmark{7},
R.~C.~Smith\altaffilmark{10},
F.~Sobreira\altaffilmark{18,35},
E.~Suchyta\altaffilmark{36},
G.~Tarle\altaffilmark{34},
D.~Thomas\altaffilmark{24},
D.~L.~Tucker\altaffilmark{1},
A.~R.~Walker\altaffilmark{10},
R.~H.~Wechsler\altaffilmark{37,16,17},
W.~Wester\altaffilmark{1},
B.~Yanny\altaffilmark{1}
\\ \vspace{0.2cm} (DES Collaboration) \\
}

\affil{$^{1}$ Fermi National Accelerator Laboratory, P. O. Box 500, Batavia, IL 60510, USA}
\affil{$^{2}$ George P. and Cynthia Woods Mitchell Institute for Fundamental Physics and Astronomy, and Department of Physics and Astronomy, Texas A\&M University, College Station, TX 77843,  USA}
\affil{$^{3}$ Observatories of the Carnegie Institution of Washington, 813 Santa Barbara St., Pasadena, CA 91101, USA}
\affil{$^{4}$ LSST, 933 North Cherry Avenue, Tucson, AZ 85721, USA}
\affil{$^{5}$ Wisconsin IceCube Particle Astrophysics Center (WIPAC), Madison, WI 53703, USA}
\affil{$^{6}$ Department of Physics, University of Wisconsin-Madison, Madison, WI 53706, USA}
\affil{$^{7}$ Instituto de F\'isica Te\'orica UAM/CSIC, Universidad Aut\'onoma de Madrid, Cantoblanco, 28049 Madrid, Spain}
\affil{$^{8}$ Kavli Institute for Cosmological Physics, University of Chicago, Chicago, IL 60637, USA}
\affil{$^{9}$ Astronomy Department, University of Washington, Box 351580, Seattle, WA 98195, USA}
\affil{$^{10}$ Cerro Tololo Inter-American Observatory, National Optical Astronomy Observatory, Casilla 603, La Serena, Chile}
\affil{$^{11}$ Department of Physics, University of Surrey, Guildford GU2 7XH, UK}
\affil{$^{12}$ CNRS, UMR 7095, Institut d'Astrophysique de Paris, F-75014, Paris, France}
\affil{$^{13}$ Department of Physics \& Astronomy, University College London, Gower Street, London, WC1E 6BT, UK}
\affil{$^{14}$ Sorbonne Universit\'es, UPMC Univ Paris 06, UMR 7095, Institut d'Astrophysique de Paris, F-75014, Paris, France}
\affil{$^{15}$ Department of Physics and Astronomy, University of Pennsylvania, Philadelphia, PA 19104, USA}
\affil{$^{16}$ Kavli Institute for Particle Astrophysics \& Cosmology, P. O. Box 2450, Stanford University, Stanford, CA 94305, USA}
\affil{$^{17}$ SLAC National Accelerator Laboratory, Menlo Park, CA 94025, USA}
\affil{$^{18}$ Laborat\'orio Interinstitucional de e-Astronomia - LIneA, Rua Gal. Jos\'e Cristino 77, Rio de Janeiro, RJ - 20921-400, Brazil}
\affil{$^{19}$ Observat\'orio Nacional, Rua Gal. Jos\'e Cristino 77, Rio de Janeiro, RJ - 20921-400, Brazil}
\affil{$^{20}$ Department of Astronomy, University of Illinois, 1002 W. Green Street, Urbana, IL 61801, USA}
\affil{$^{21}$ National Center for Supercomputing Applications, 1205 West Clark St., Urbana, IL 61801, USA}
\affil{$^{22}$ Institut de Ci\`encies de l'Espai, IEEC-CSIC, Campus UAB, Carrer de Can Magrans, s/n,  08193 Bellaterra, Barcelona, Spain}
\affil{$^{23}$ Institut de F\'{\i}sica d'Altes Energies (IFAE), The Barcelona Institute of Science and Technology, Campus UAB, 08193 Bellaterra (Barcelona) Spain}
\affil{$^{24}$ Institute of Cosmology \& Gravitation, University of Portsmouth, Portsmouth, PO1 3FX, UK}
\affil{$^{25}$ School of Physics and Astronomy, University of Southampton,  Southampton, SO17 1BJ, UK}
\affil{$^{26}$ Department of Physics, IIT Hyderabad, Kandi, Telangana 502285, India}
\affil{$^{27}$ Jet Propulsion Laboratory, California Institute of Technology, 4800 Oak Grove Dr., Pasadena, CA 91109, USA}
\affil{$^{28}$ Department of Astronomy, University of California, Berkeley,  501 Campbell Hall, Berkeley, CA 94720, USA}
\affil{$^{29}$ Lawrence Berkeley National Laboratory, 1 Cyclotron Road, Berkeley, CA 94720, USA}
\affil{$^{30}$ Australian Astronomical Observatory, North Ryde, NSW 2113, Australia}
\affil{$^{31}$ Instituci\'o Catalana de Recerca i Estudis Avan\c{c}ats, E-08010 Barcelona, Spain}
\affil{$^{32}$ Department of Physics and Astronomy, Pevensey Building, University of Sussex, Brighton, BN1 9QH, UK}
\affil{$^{33}$ Instituto de F\'\i sica, UFRGS, Caixa Postal 15051, Porto Alegre, RS - 91501-970, Brazil}
\affil{$^{34}$ Department of Physics, University of Michigan, Ann Arbor, MI 48109, USA}
\affil{$^{35}$ Universidade Federal do ABC, Centro de Ci\^encias Naturais e Humanas, Av. dos Estados, 5001, Santo Andr\'e, SP, Brazil, 09210-580}
\affil{$^{36}$ Computer Science and Mathematics Division, Oak Ridge National Laboratory, Oak Ridge, TN 37831}
\affil{$^{37}$ Department of Physics, Stanford University, 382 Via Pueblo Mall, Stanford, CA 94305, USA}

\begin{abstract}
We present Magellan/IMACS spectroscopy of the recently-discovered Milky Way satellite Eridanus~II (Eri~II). We identify 28 member stars in Eri~II, from which we measure a systemic radial velocity of $v_{\rm hel} = 75.6 \pm 1.3~\mbox{(stat.)} \pm 2.0~\mbox{(sys.)}$~\kms and a velocity dispersion of $6.9^{+1.2}_{-0.9}$~\kms. Assuming that Eri~II is a dispersion-supported system in dynamical equilibrium, we derive a mass within the half-light radius of $1.2^{+0.4}_{-0.3} \times 10^{7}$~\msun, indicating a mass-to-light ratio of $420^{+210}_{-140}$~\msun/\lsun and confirming that it is a dark matter-dominated dwarf galaxy. From the equivalent width measurements of the CaT lines of 16 red giant member stars, we derive a mean metallicity of $\feh = -2.38 \pm 0.13$ and a metallicity dispersion of $\sigma_{\feh} = 0.47 ^{+0.12}_{-0.09}$. The velocity of Eri~II in the Galactic Standard of Rest frame is $v_{\rm GSR} = -66.6$~\kms, indicating that either Eri~II is falling into the Milky Way potential for the first time or it has passed the apocenter of its orbit on a subsequent passage. At a Galactocentric distance of $\sim$370 kpc, Eri II is one of the Milky Way's most distant satellites known. Additionally, we show that the bright blue stars previously suggested to be a young stellar population are not associated with Eri~II.  The lack of gas and recent star formation in Eri~II is surprising given its mass and distance from the Milky Way, and may place constraints on models of quenching in dwarf galaxies and on the distribution of hot gas in the Milky Way halo. Furthermore, the large  velocity dispersion of Eri II can be combined with the existence of a central star cluster to constrain MACHO dark matter with mass $\gtrsim$10~\msun. 
\end{abstract}


\section{INTRODUCTION}
\label{intro}

Over the past two years, more than 20 ultra-faint dwarf galaxy candidates have been discovered in data from the Dark Energy Survey~\citep[DES;][]{bechtol15,koposov15,dw15b,kim15_hor2} and other large optical surveys~\citep{martin15,laevens15,laevens15b,kim15_peg3,kim15_kim2,torrealba16,torrealba16b,Homma2016,dw16}. One of the largest, most luminous, and most distant newly discovered satellites is Eridanus~II (Eri II), which has an absolute magnitude of $M_V \sim -7$, a half-light radius of $r_h\sim280$~pc, and a Galactocentric distance of $D \sim370$~kpc~\citep{bechtol15,koposov15,Crnojevic16}.

Eri II is likely located just beyond the virial radius of the Milky Way, which is typically estimated to be $\sim$300~\kpc~\citep{Taylor2016, Bland-Hawthorn2016}. 
This places Eri II in a sharp transition region between the gas-free dwarf spheroidals (with $D\lesssim250$~kpc) and the more distant gas-rich star-forming dwarfs~\citep{Einasto1974,Blitz2000,Grcevich2009,Spekkens2014}. 
\citet{koposov15} suggested that Eri II may contain a young stellar population component ($\sim$250 Myr) due to the spatial coincidence of a few candidate blue loop stars, and therefore Eri~II would be similar to the gas-rich dwarf Leo T, which is slightly more distant and more luminous~\citep[$D=420$~kpc, $M_V \sim -8$;][]{Irwin2007,deJong2008,Ryan-Weber2008} and has undergone multiple epochs of star formation~\citep{deJong2008, Weisz2012}. If this were the case, Eri~II would be the least luminous star-forming galaxy known. However, \citet{Westmeier2015} measured the \ion{H}{1} gas content using HIPASS data~\citep{Barnes2001}, and did not detect any \ion{H}{1} gas associated with Eri~II. \citet{Crnojevic16} also obtained the \ion{H}{1} observations from the Green Bank Telescope, and found a more stringent upper limit on the \ion{H}{1} mass ($M_{\rm HI} < 2800~\msun$), indicating that Eri~II is an extremely gas-poor system. 
With deep imaging from Magellan/Megacam, \citet{Crnojevic16} found a possible intermediate-age ($\sim$3 Gyr) population in Eri~II. Moreover, they confirmed that there is a star cluster whose projected position is very close to the center of Eri~II, making it the least luminous galaxy known to host a (possibly central) star cluster.

The features described above make Eri~II one of the most interesting of the newly discovered Milky Way satellites for spectroscopic study. In this paper we present the first spectroscopic observations of Eri II, from which we determine its dark matter content and test for the existence of a young stellar population. In \S\ref{observations} we describe the observations, target selection, and data reduction.  In \S\ref{measurements} we perform velocity and metallicity measurements on the observed stars in Eri~II and identify spectroscopic members. We then compute the global properties of Eri~II and discuss its nature and origin in \S\ref{discussion}, and conclude in \S\ref{conclusions}.

\section{OBSERVATIONS AND DATA REDUCTION}
\label{observations}

\subsection{Observations and Target Selection}
\label{sec:targets}

We observed Eri~II with the IMACS spectrograph~\citep{dressler06} on the Magellan Baade telescope on the nights of 2015~October~16-17 and 2015~November~18--19.
The observing conditions on both runs were clear, with seeing that varied from 0\farcs6 to 0\farcs9. The observing procedure and instrument setup are similar to the observation of the Tucana~III (Tuc~III) dwarf galaxy described by~\citet{tuc3}. For both runs we used the f/4 camera on IMACS, which has an $8192 \times 8192$ mosaic provided by a $4 \times 2$ array of $2048 \times 4096$ pixel e2v CCDs. The spectrograph was configured with the 1200~$\ell$/mm grating blazed at 32.7\degr. This setup produces a spectral dispersion of 0.19~\AA~pix$^{-1}$, a peak throughput above 14\%\ for $7800-8800$~\AA, and a spectral resolution of $R\sim11,000$ for a 0\farcs7 slit width. We used a tilt angle of 32.4\degr\ to provide a minimum wavelength range of $7550-8750$~\AA\ for each slit, with typical wavelength coverage of $7400-9000$~\AA.  The WB5600-9200 filter was used to block shorter wavelength light. This wavelength range covers the Ca triplet (CaT) absorption lines around 8500~\AA\, used for measuring radial velocities and metallicities of candidate member stars, as well as the telluric absorption lines (Fraunhofer A-band) around 7600~\AA\, used for the corrections of velocity errors caused by mis-centering of the stars within the slits (see~\S\ref{RV} and \citealt{sohn07} for details). While the f/4 camera on IMACS provides a full field-of-view of $15.4\arcmin \times 15.4\arcmin$ for multi-slit spectroscopy, we limited the placement of slits to a $15.4\arcmin \times 8\arcmin$ portion to ensure that all the spectra span the required wavelength range for accurate velocity measurements.

We observed the candidate member stars in Eri~II with one slitmask. The spectroscopic targets were selected using photometry from the coadded images of the first internal annual release of DES data~\citep[Y1A1;][]{bechtol15}. Since Eri~II is a distant Milky Way satellite, most of the candidate stars brighter than $g=23$ are near the tip of the red giant branch (RGB). We chose spectroscopic targets using a PARSEC isochrone \citep{bressan12} with age~$= 12$~Gyr and $\feh = -2.2$ as guidance.  RGB candidates were selected as stars within 0.13~mag of the isochrone, brighter than $g = 22.5$, and within $7\arcmin$ (three times the half-light radius) of the center of Eri~II.  
In addition, we targeted potential blue loop stars within a box defined by $0.2 < g-r < 0.4$ and $20.3 < g < 20.6$.  The relative priorities for RGB stars were based on brightness and projected distance from the center of the galaxy, and for blue loop candidates the priorities were based only on projected distance on the sky (since all of the stars have similar magnitudes).
Remaining mask space was filled with stars that have photometry that makes them unlikely to be members. This selection process resulted in the placement of 68 $0\farcs7 \times 5.0\arcsec$ slitlets on the slitmask.
We observed this mask for a total of 3~hrs on the October run and 9~hrs on the November run. To ensure accurate velocity measurements, after every two 30-40 min science exposures, we acquired one wavelength calibration frame and one flatfield frame at the same position as the science exposure. For the October run, we used He, Ne, and Ar comparison lamps for wavelength calibration, while for the November run, we used Kr, Ne, and Ar lamps. The Kr lamp provides additional strong lines in the critical $7600 - 7900$~\AA\ wavelength range where there are few usable Ne and Ar lines, improving the wavelength calibration around the Fraunhofer A-band.

In addition to the observations targeting candidate members of Eri~II, we also obtained spectra of several metal-poor stars to serve as radial velocity templates for the velocity measurements, and a hot, rapid rotator (HR~4781) to serve as a telluric template for the velocity error corrections. More templates were also obtained during additional IMACS runs with identical observing setups. For both the radial velocity templates and the telluric template, we obtained the spectra using a north-south oriented longslit while driving the stars perpendicularly across the slit (i.e., across the $0\farcs7$ dimension) at a constant rate during the exposure. These spectra simulate a source that uniformly fills the slit, and thus accurately reflect the mean integrated slit function.

\subsection{Data Reduction}
\label{sec:reductions}

We reduced the IMACS spectra following the procedures described by~\citet{tuc3} for Tuc~III. The reduction procedures include bias subtraction, removal of read-out pattern noise, an initial wavelength solution and slit mapping with the Cosmos pipeline~\citep{dressler11}, and a refined wavelength calibration and spectral extraction using an IMACS pipeline derived from the DEEP2 data reduction pipeline for Keck/DEIMOS \citep{cooper12}. 

\begin{figure*}[th!]
\epsscale{1}
\plotone{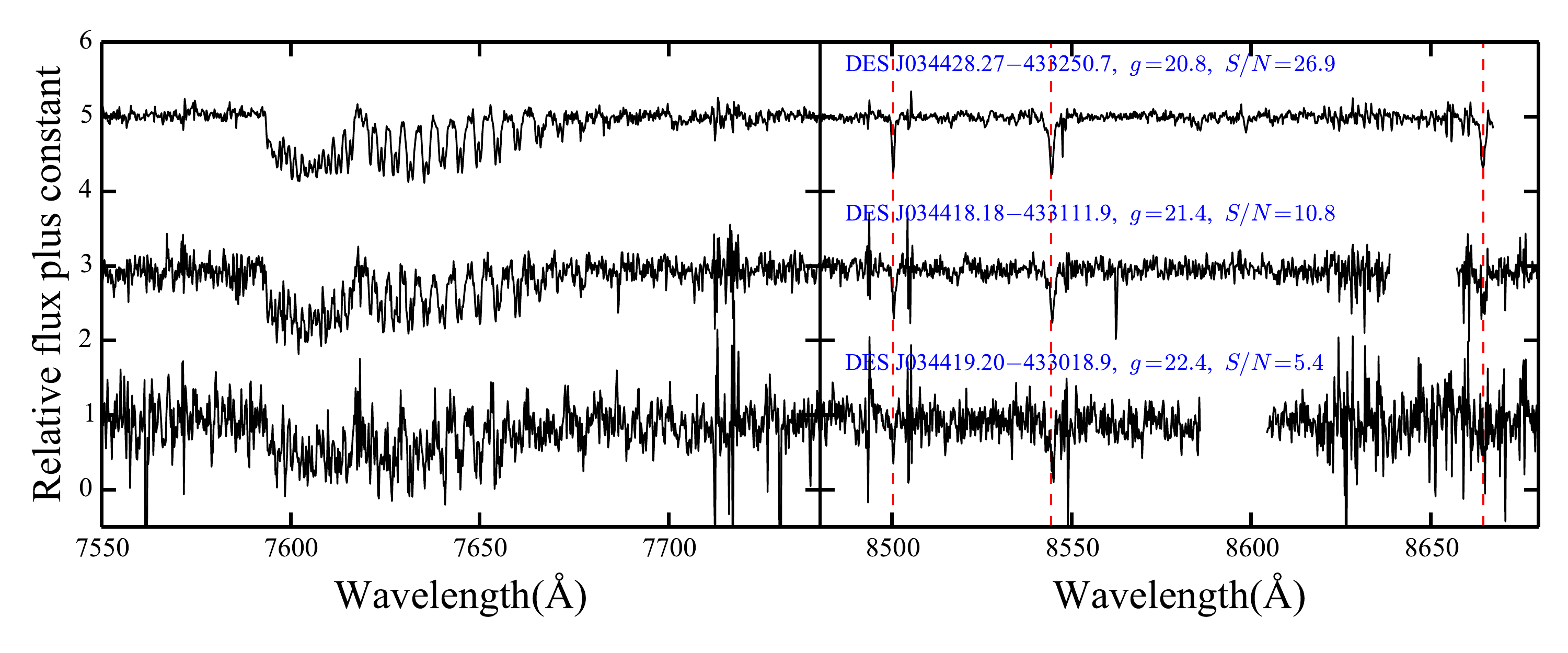}
\caption{IMACS spectra of three Eri~II member stars at various magnitudes and S/N levels. \emph{(Left)} The Fraunhofer A-band region of the spectrum, used for the corrections of velocity errors caused by the mis-centering of the stars within the slits. \emph{(Right)} The CaT region, used for measuring radial velocities and metallicities of candidate member stars. The three Ca lines are marked with dashed red lines. The gaps in the spectra are caused by the gaps between IMACS CCDs.  
}
\label{spectra}
\end{figure*}

Each individual science frame was reduced using the corresponding flatfield frame and wavelength calibration frame. The end products of the pipeline are the extracted 1D spectra and the corresponding inverse-variance spectra. For the targeted 68 stars, 66 were successfully extracted, and 2 stars fell onto chip gaps or off the detector array.
We then combined the extracted spectra from each observing run using inverse-variance weighting. As the November run has a much longer total exposure time and better seeing, we used the coadded spectra from the November run for the kinematic measurements in later sections. We kept the coadded spectra from the October run separate from the  coadded spectra from the November run to test the possible radial velocity variation from binary orbital motion. For the November coadded spectra, we reached signal-to-noise ratio (S/N) $\sim$5 per pixel for stars at g$\sim$22 and S/N$\sim$30 per pixel for stars at g$\sim$20.5. For the October coadded spectra, the S/N is about a factor of two lower.
Finally, all the coadded spectra were normalized to unity in the continuum by fitting a second-order polynomial. Examples of the spectra at various brightness and S/N levels are shown in Figure~\ref{spectra}.

We reduced the spectra for the velocity and telluric templates in the same manner as the science exposures described above. For the telluric template, we set the regions outside of telluric absorption bands to unity; for the velocity templates, we set the regions inside the telluric bands to unity and shifted them to rest frame.

\section{VELOCITY AND METALLICITY MEASUREMENTS}
\label{measurements}
\subsection{Radial Velocity Measurements}\label{RV}

\setcounter{footnote}{0}

We measured radial velocities by fitting reduced spectra with velocity templates using a Markov Chain Monte Carlo (MCMC) sampler \citep[\code{emcee};][]{emcee},\footnote{\code{emcee} v2.2.0: http://dan.iel.fm/emcee/} and a likelihood function defined as:


\begin{equation}\label{mle}
\small
\log \mathcal{L} = -\frac{1}{2}\sum_{\lambda = \lambda_1}^{\lambda_2} \frac{[f_{s}(\lambda) - f_{\rm std}\big(\lambda(1 + \frac{v}{c})\big)]^2}{\sigma_{s}^2(\lambda)}.
\end{equation}

\noindent
Note that the log-likelihood is defined up to an additive constant. Here, $f_{s}(\lambda)$ and $\sigma_{s}^2(\lambda)$ are the normalized spectrum and its corresponding variance, while $f_{\rm std}(\lambda)$ is the normalized velocity template. For the velocity measurement, we primarily used the CaT feature, and therefore set $\lambda_1 = 8450$~\AA\ and $\lambda_2 = 8700$~\AA.
Our procedure fits the radial velocity by shifting the velocity template by a velocity $v$ to maximize the likelihood function. For this paper, we use the metal-poor RGB star HD~122563 as the template for all of the science spectra.

For each spectrum, we ran an MCMC sampler with 20 walkers that each made 1000 steps including a burn-in stage of 50 steps. We used the median and the standard deviation (with $5\sigma$ clipping) of the posterior distributions as the measured velocity~$v_{\rm obs}$ and velocity error~$\sigma_{v_{\rm obs}}$ for each star.

We then determined and applied a telluric correction to each velocity measurement to account for velocity errors that result from mis-centering the star within the slit. We ran the same MCMC sampler as described above, but instead used a telluric template and a fitting wavelength range of $7550-7700$~\AA. For each spectrum, the telluric correction $v_{\rm tel}$ and uncertainty $\sigma_{v_{\rm tel}}$ were obtained from the posterior distribution from the MCMC sampler. 

\begin{figure*}[th!]
\epsscale{1}
\plotone{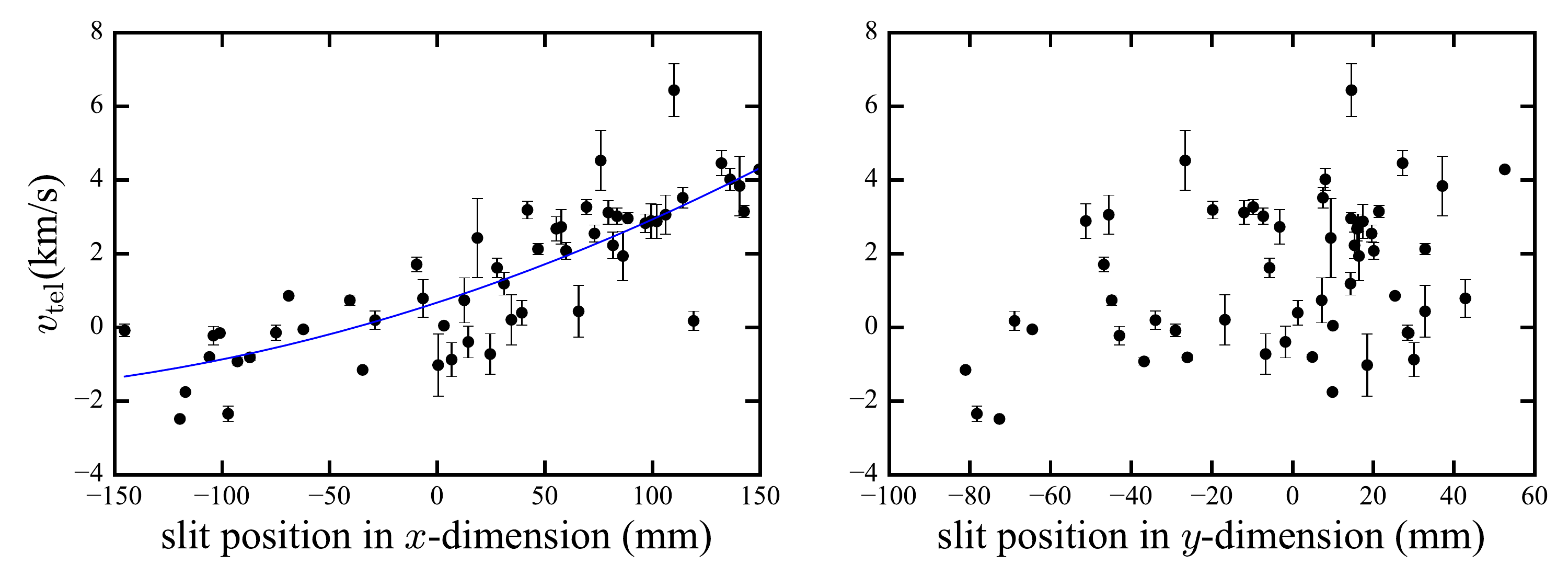}
\caption{The telluric correction $v_{\rm tel}$ as a function of slit position in the direction parallel to the slits ($x$-dimension, \emph{left panel}) and perpendicular to the slits ($y$-dimension, \emph{right panel}). The RMS of $v_{\rm tel}$ residuals in the $x$-dimension with respect to a fitted second-order polynomial is $\sim$1~\kms. 
}
\label{tel_corr}
\end{figure*}

In Figure~\ref{tel_corr}, we show the telluric correction $v_{\rm tel}$ as a function of slit position in the direction parallel to the slits ($x$) and perpendicular to the slits ($y$). The correction generally ranges between $-$2~\kms and 4~\kms for this mask and shows a dependence on the slit position in the spatial direction.
The RMS of the residuals after a second-order polynomial fit to the data is $\sim$1~\kms. 
The systematic trend (i.e., polynomial fit) of this telluric correction is likely caused by either a small mask rotation or anamorphic demagnification of the IMACS spectrograph; the scatter around the fit (i.e., RMS) may be associated with the astrometric errors of DES Y1A1 data and the systematic uncertainty in the velocity correction determination.

The velocity was then calculated as $v = v_{\rm obs} - v_{\rm tel}$ and the statistical uncertainty of $v$ was calculated as $\sigma_{v_{\rm stat}} = \sqrt{\sigma^2_{v_{\rm obs}}+\sigma^2_{v_{\rm tel}}}$. Note that we used $v_{\rm tel}$ from individual stars for the telluric correction rather than the polynomial fit shown in Figure~\ref{tel_corr}.
It is worth noting that $\sigma_{v_{\rm stat}}$ is only the statistical uncertainty on the velocity measurements, which is associated with the S/N of the spectra. For high S/N spectra, the velocity measurement can be very precise. However, the accuracy of the velocity measurement is limited by systematic effects, such as instrument flexure, uncertainties in the wavelength calibration, uncertainties in the template velocity, template mismatching, and the uncertainties in the telluric correction. These systematic uncertainties should also be considered in the total error budget. We estimated the systematic uncertainty as the quadrature difference between repeat measurements and the statistical uncertainty~\citep[c.f.][]{sg07,simon15b}. Similar to~\citet{tuc3}, we found that this systematic uncertainty is $\sigma_{v_{\rm sys}}=1.2$~\kms for the October observations and $\sigma_{v_{\rm sys}}=1.0$~\kms for the November observations. The slight difference in the systematic errors between the two observing runs is mainly because the new Kr lamp included in November improved the wavelength solution at the blue end. We added this systematic uncertainty in quadrature with the statistical uncertainties as the final reported velocity uncertainties, $\sigma_v = \sqrt{\sigma^2_{v_{\rm stat}}+\sigma^2_{v_{\rm sys}}}$. 

Of the 66 extracted spectra, 54 have high enough S/N to determine velocities and velocity uncertainties using the aforementioned method. Finally, all velocity measurements are transformed to the heliocentric frame. The results are listed in Table~\ref{tab:eri2_spec}.

\begin{figure*}[th!]
\epsscale{1}
\plotone{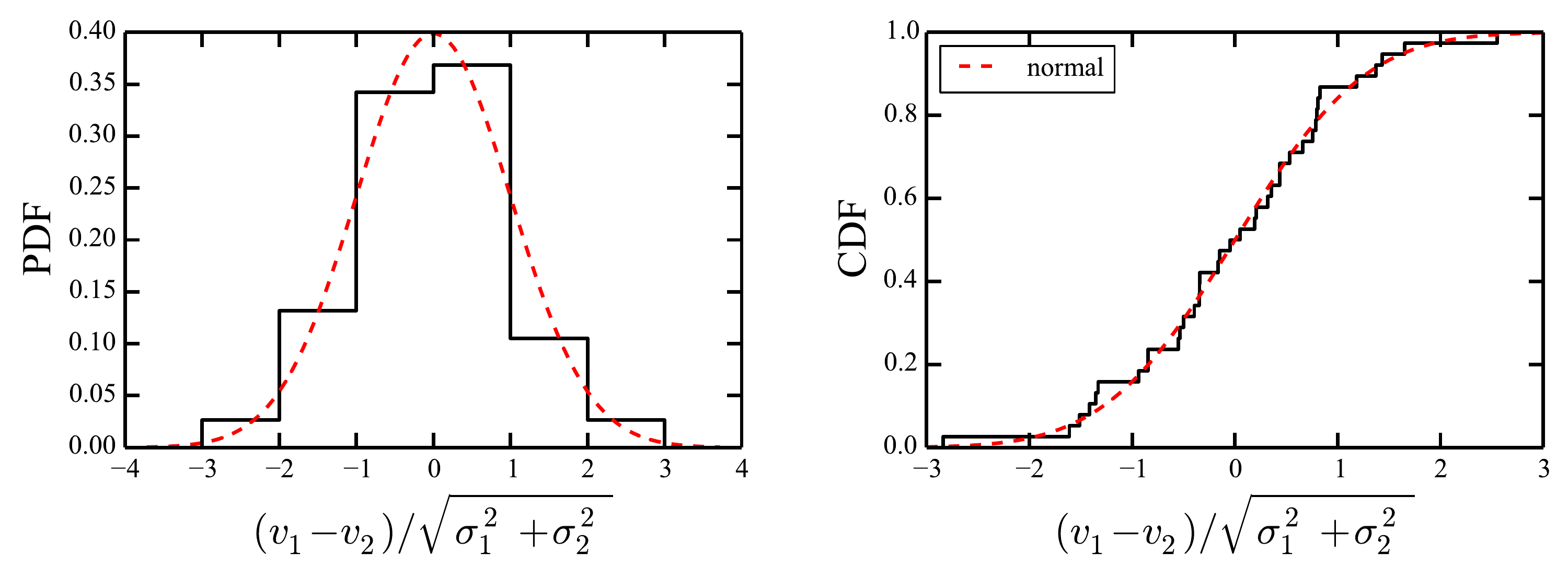}
\caption{The results of radial velocity uncertainty estimation tests using 38 pairs of repeated observations from the October run ($v_1$, $\sigma_1$) and November run ($v_2$, $\sigma_2$). The probability distribution function (PDF, \emph{left panel}) and cumulative distribution function (CDF, \emph{right panel}) show the distributions of the velocity difference normalized by the quadrature sum of their uncertainties. The red dashed curves show a standard normal distribution with zero mean and unit variance. The p-value from a K-S test between the sample and the model is 0.98. This indicates that our estimation of the velocity uncertainties is reasonable.
}
\label{v_err}
\end{figure*}

In order to confirm that our error estimation is reasonable, we select 38 stars that have measured velocities and velocity uncertainties from both the November run and October run, and compute the distribution of velocity differences between the two independent measurements ($v_1$, $v_2$), divided by the quadrature sum of their uncertainties ($\sqrt{\sigma_1^2+\sigma_2^2}$); $v_1$ and $\sigma_1$ are the measurements from October and $v_2$ and $\sigma_2$ from November. 
The resulting distribution shown in Figure~\ref{v_err} is well-described by a normal distribution with zero mean and unit variance shown as a red dashed curve in the same plot. A Kolmogorov-Smirnov (K-S) test of the repeated measurements against the normal distribution gives a p-value of 0.98, confirming that our error model provides an accurate description of the velocity uncertainties. Since we do not see any outliers when comparing the results between the two observing runs, we conclude that we are not able to detect any binary stars in Eri~II based on the one month baseline. 
Binaries with longer periods might be present, but detecting them would require additional observations with a longer time baseline.

\subsection{Spectroscopic Membership Determination}
\label{membership}

\begin{figure*}[th!]
\epsscale{1.2}
\plotone{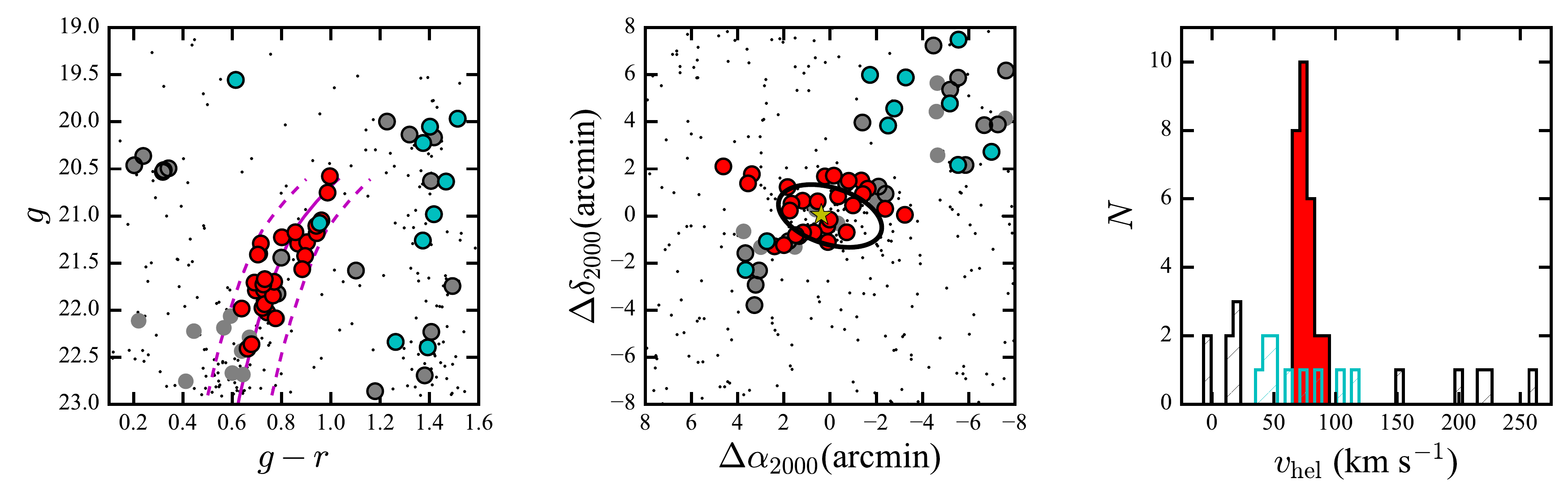}
\caption{\emph{(Left)} Color-magnitude diagram of Eri~II using DES Y1A1 photometry.  Stars within 8\arcmin\ of the center of Eri~II are plotted as small black dots, and stars selected for spectroscopy (as described in \S\ref{sec:targets}) are plotted as filled gray circles.  Points surrounded by black outlines represent the stars for which we obtained successful velocity measurements. Those we identify as Eri~II members are filled in with red. Non-members that have velocities close to the velocity of Eri~II are filled in with cyan. A PARSEC isochrone \citep{bressan12} with age = 12.0 Gyr and $\feh = -2.2$ is displayed as the solid magenta line. The other two dashed magenta lines show the boundaries of the selected high priority RGB candidates as discussed in~\S\ref{sec:targets}. \emph{(Middle)} Spatial distribution of the observed stars. Symbols are as in the left panel.  The elliptical half-light radius of Eri~II is outlined as a black ellipse.
The yellow star indicates the location of the central star cluster of Eri~II. \emph{(Right)} Radial velocity distribution of observed stars. The clear narrow peak of stars at $v \sim 75$~\kms highlighted in red is the signature of Eri~II. The hatched histogram indicates stars that are non-members of Eri~II, among which the hatched cyan histogram corresponds to the cyan filled circles in the left and middle panels.}
\label{cmd}
\end{figure*}

\begin{figure}[th!]
\epsscale{1.1}
\plotone{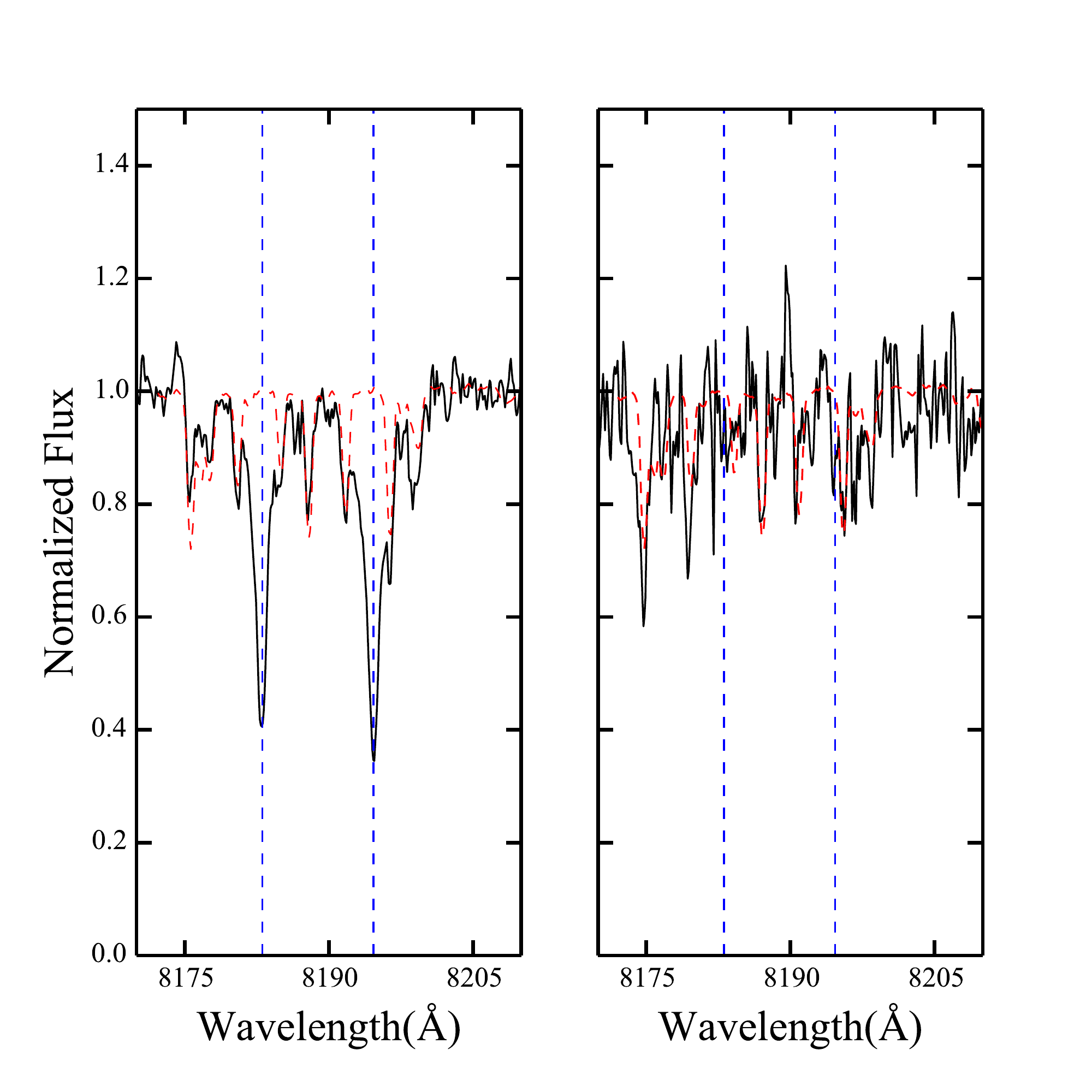}
\caption{
Example of rest frame spectra of a foreground main sequence star (\emph{left}) and an Eri~II member star (\emph{right}) around the \ion{Na}{1}~$\lambda$8190~\AA\ doublet, shown in black. The two \ion{Na}{1} lines are marked with dashed blue vertical lines. The foreground dwarf has strong \ion{Na}{1} lines, while the Eri~II member is a giant star with low surface gravity and therefore the \ion{Na}{1} lines are hardly detectable. Overplotted dashed red lines are the spectrum of the telluric standard star, indicating the absorption from the Earth's atmosphere.
}
\label{nai}
\end{figure}

The color-magnitude diagram (CMD), spatial distribution, and velocity distribution of the candidate stars are displayed in Figure~\ref{cmd}. From the 54 stars with measured velocities, we found 28 Eri~II members that form a narrow velocity peak at $\sim$75~\kms (right panel of Figure~\ref{cmd}).
For the large majority of the observed stars, membership status is unambiguous. 
The member stars located close to the center of Eri~II and located along the isochrone in the CMD are highlighted in red in the histogram and denoted as red filled circles.  
Stars with $v_{\rm hel} > 140~\kms$ or $v_{\rm hel} < 30 ~\kms$ that are clearly not associated with Eri~II are shown by gray filled circles in the left and middle panels of Figure~\ref{cmd}. 
Several candidate members have a velocity close to the peak and are highlighted in cyan in the histogram, and also denoted as cyan filled circles in the left and middle panels of Figure~\ref{cmd}. 
These stars are classified as non-members since they are located far from the RGB isochrone in the CMD.  
While DES~J034404.78$-$432727.7 lies close to the RGB isochrone, it is more than two half-light radii away from the center of Eri~II along the minor axis and has strong \ion{Na}{1} lines. In fact, all the stars coded in cyan show strong \ion{Na}{1} lines at $\lambda = 8183$~\AA~and $\lambda = 8195$~\AA~(see Figure~\ref{nai} as an example), indicating that they are foreground M-dwarf stars rather than distant giants~\citep{Schiavon1997}. 

Finally, we use the Besan\c{c}on~\citep{besancon} Galactic stellar model to estimate the expected number of foreground main sequence stars in our spectroscopic sample.  We select simulated stars within 0.2 mag of the PARSEC isochrone and with $20.5 < g < 22.5$ (i.e., the location of the red filled circles in the CMD in the left panel of Figure~\ref{cmd}). We found $\sim$70 simulated stars that have a velocity consistent with the heliocentric velocity peak of Eri II (60 -- 90~\kms) in an area of 1~deg$^2$ centered on Eri~II. When scaled to the area within two half-light radii of Eri~II ($\sim$0.02 deg$^2$) the foreground contamination is expected to be 1--2 stars.

\subsection{Metallicity Measurements}


We measured the metallicity of the red giant members using the equivalent widths (EWs) of the CaT lines. Following the procedure described by~\citet{simon15b} and \citet{tuc3}, we fit all three of the CaT lines with a Gaussian plus Lorentzian function and then converted the summed EWs of the three CaT lines to metallicity using the calibration relation from~\citet{carrera13}.
Because the horizontal branch stars of Eri~II are too faint for accurate measurements in DES imaging, we used the absolute $V$ magnitude for the CaT calibration. We first performed the color-transformation from DES-$g$ and DES-$r$ to apparent $V$ magnitude using Equation (5) in~\citet{bechtol15} and then adopted the distance modulus $(m-M) = 22.8$ derived by~\citet{Crnojevic16} to calculate absolute magnitudes. 

Among the 28 spectroscopic members determined in \S\ref{membership}, 16 of them have successful metallicity measurements. The other members either do not have large enough S/N for EW measurements, or have one of the three Ca lines falling onto the CCD chip gap. The measured metallicities are reported in Table~\ref{tab:eri2_spec}. 

The statistical uncertainties on the EWs are calculated from the Gaussian and Lorentzian fit. We then compute a systematic uncertainty of 0.2~\AA~on the summed EWs derived with repeat measurements (using the same approach as for the systematic velocity uncertainty in \S\ref{RV}). The final uncertainties on the EWs reported in  Table~\ref{tab:eri2_spec} are the quadrature sum of the statistical and systematic uncertainties.
The metallicity uncertainties shown in Table~\ref{tab:eri2_spec} are dominated by the uncertainties of the CaT EWs, with small contributions from the uncertainties of the distance modulus, the stellar photometry, and the uncertainties on the calibration parameters from~\citet{carrera13}.

\begin{deluxetable}{llr}
\tablecaption{Summary of Properties of Eridanus\,II}
\tablewidth{0pt}
\tablehead{
\colhead{Row} & \colhead{Quantity} & \colhead{Value}
}
\startdata
(1) & RA (J2000)                            & 03:44:20.1 \\
(2) & Dec (J2000)                           & $-$43:32:01.7 \\
(3) & Heliocentric Distance (kpc)           & $366  \pm 17$  \\
(4) & Galactocentric Distance (kpc)         & $368  \pm 17$  \\
(5) & $M_{V,0}$                             & $-7.1 \pm 0.3$  \\
(6) & $L_{V,0}$ (\lsun)                     & $5.9^{+1.9}_{-1.4} \times 10^4$ \\
(7) & $r_{\rm 1/2}$ (pc)                    & $277 \pm 14$  \\
(8) & $r_{\rm 1/2}$ (arcmin)                & $2.31 \pm 0.12$ \\
(9) & $\epsilon$                            & $0.48 \pm 0.04$ \\
(10) & PA (N to E; deg)                      & $72.6 \pm 3.3$ \\ [0.5em]
\hline\\[-0.5em]
(11)  & $v_{\rm hel}$ (\kms)                & $75.6 \pm 1.3 \pm 2.0$  \\
(12)  & $v_{\rm GSR}$ (\kms)                & \vgsr    \\
(13)  & $\sigma_v$ (\kms)                     & \vdisp \\
(14)  & $M_{\rm half}$ (\msun)              & \mass  \\
(15)  & $M/L_{V}$ (\msun/\lsun)             & \masstolight \\
(16)  & $\frac{dv}{d\chi}$ (\kms~arcmin$^{-1}$) & $0.1 \pm 1.1$\\
(17)  & Mean metallicity                         & \fehmedian \\
(18)  & Metallicity dispersion (dex)        & \fehdisp \\
(19)  & $\log_{10}{J(0.2\degr)}$ (GeV$^{2}$~cm$^{-5}$) & \jsmall \\
(20)  & $\log_{10}{J(0.5\degr)}$ (GeV$^{2}$~cm$^{-5}$) & \jlarge \\[-0.5em]
\enddata

\tablecomments{Rows (1)-(10) are taken or derived from \citet{Crnojevic16}.  Values in rows (11)-(20) are derived using the measurements in this paper. All values reported here (and in this paper) are from the 50th percentile of the posterior probability distributions. The uncertainties are from the 16th and 84th percentiles of the posterior probability distributions.}
\label{eri2_table}
\end{deluxetable}

\section{DISCUSSION}
\label{discussion}

In this section we determine the global properties of Eri~II and discuss its nature and origin. We then consider implications for the quenching of star formation in dwarf galaxies and constraints on the nature of dark matter.

\subsection{Velocity Dispersion and Mass}
\label{sec:kinematics}

\begin{figure*}[th!]
\epsscale{1}
\plottwo{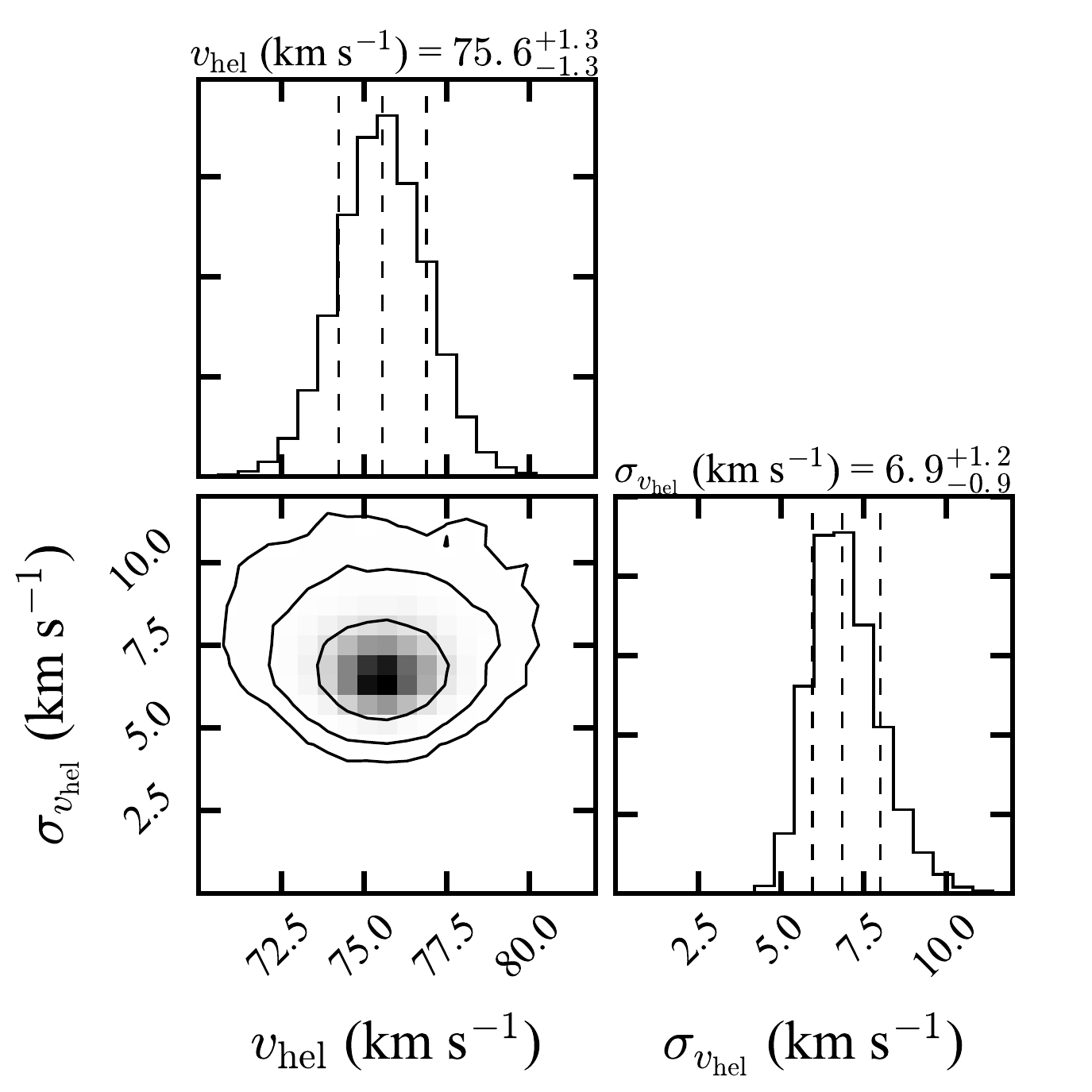}{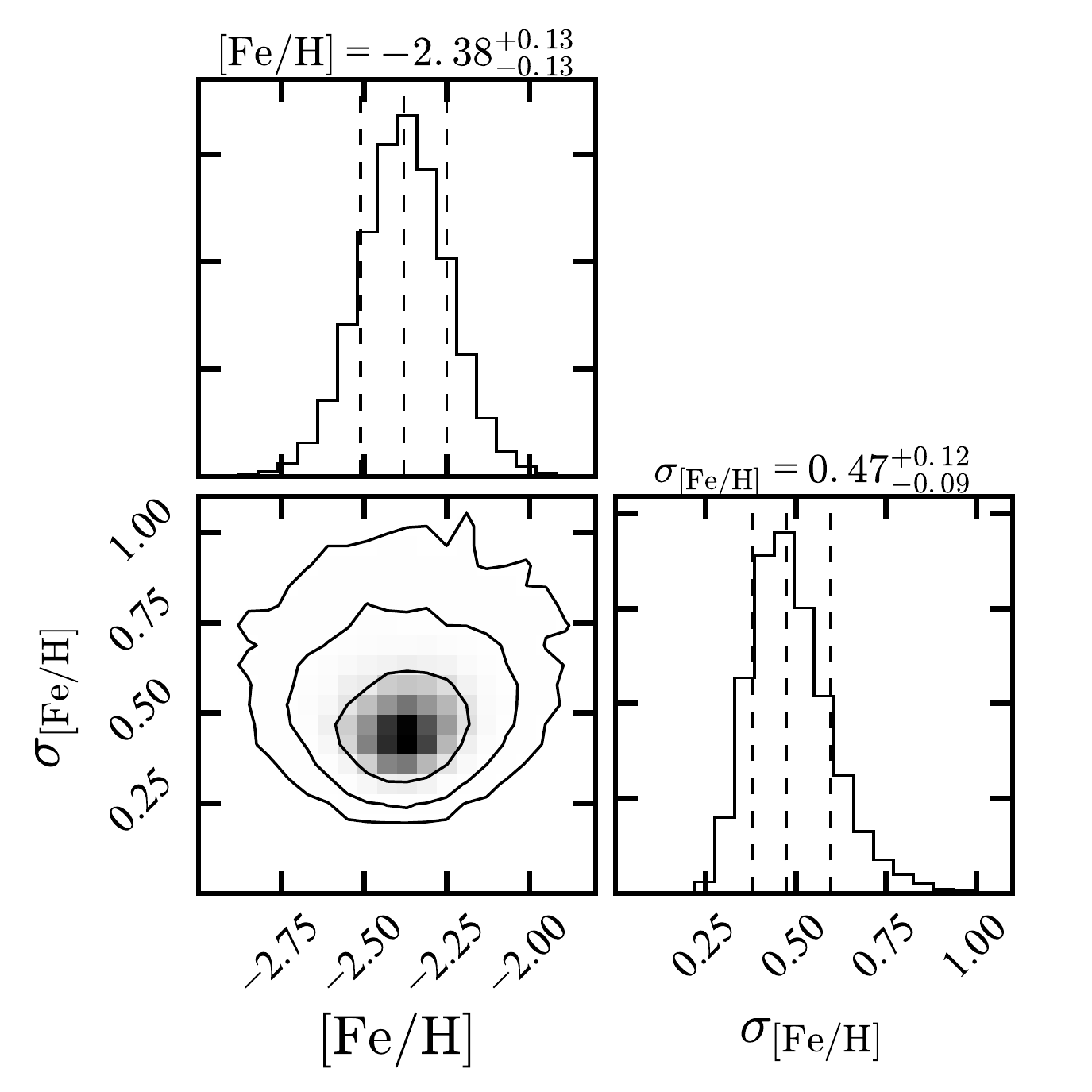}
\caption{Two-dimensional and marginalized posterior probability distribution from an MCMC sampler using a likelihood model for the systemic velocity and velocity dispersion \emph{(left)} and the mean metallicity and metallicity dispersion \emph{(right)} of Eri~II.  For the 1-D histograms, the 16th, 50th, and 84th percentiles are indicated by dashed lines. For the 2-D histograms, contours represent the 68\%, 95.5\%, and 99.7\% confidence intervals.}
\label{vmcmc}
\end{figure*}

With the 28 spectroscopically confirmed members, we calculated the systemic velocity and the velocity dispersion of Eri~II using a 2-parameter Gaussian likelihood function similar to that of~\citet{Walker06}:

\begin{equation}\label{mle_vdisp}
\medmuskip=-1mu
\log \mathcal{L} = -\frac{1}{2} \left [  \sum_{n = 1}^{N}\log (\sigma_{v_{\rm hel}}^2 + \sigma_{v_i}^2) +  \sum_{n = 1}^{N} \frac{(v_i - v_{\rm hel})^2}{\sigma_{v_i}^2 + \sigma_{v_{\rm hel}}^2} \right ],
\end{equation}

\noindent
where $v_{\rm hel}$ and $\sigma_{v_{\rm hel}}$ are the systemic velocity and the velocity dispersion of Eri~II, and $v_i$ and $\sigma_{v_i}$ are the velocities and velocity uncertainties for each member star as calculated in \S\ref{RV}.
We used an MCMC to sample the posterior distribution. We find a systemic velocity of $v_{\rm hel} = \vbulk$~\kms and a velocity dispersion of $\sigma_{v_{\rm hel}} = \vdisp$~\kms, where we report the median of the posterior and the uncertainty calculated from the 16th and 84th percentiles. The systematic uncertainty (2.0~\kms) on the systemic velocity is attributed to uncertainty on the velocity zero-point of the template star. The posterior probability distribution from the MCMC sampler for the kinematic properties of Eri~II is displayed on the left side of Figure~\ref{vmcmc}.

In principle, the measured velocity dispersion of Eri~II could be artificially inflated by the orbital motions of
binary stars~\citep{mc10}. As mentioned in \S\ref{RV}, our observations do not span a long enough time baseline to detect any binaries. However, studies over longer time baselines have tended to show that binary stars do not have a significant impact on the velocity dispersion of classical dwarf spheroidals \citep{olszewski96} or ultra-faint dwarfs~\citep{minor10,simon11}. Given the large velocity dispersion of Eri~II, the effect of the binaries is expected to be small and our results should be similar even if our sample contains a few binary stars.

We calculated the mass contained within the half-light radius adopting the mass estimator from \citet{wolf10}, using the velocity dispersion determined above and the half-light radius of Eri~II measured by~\citet{Crnojevic16}. The derived dynamical mass is $M_{\rm 1/2} = \mass$~\msun.  Given a luminosity within its half-light radius of $3.0^{+0.9}_{-0.7} \times 10^{4}$~\lsun, the mass-to-light ratio of Eri~II is \masstolight~\msun/\lsun. The reported uncertainties on the dynamical mass and mass-to-light ratio include the uncertainties on the velocity dispersion, half-light radius, and luminosity.

The mass estimator from \citet{wolf10} is only valid for dispersion-supported stellar systems in dynamical equilibrium. Given the distance to Eri~II, the system is very likely to be in dynamical equilibrium. Nevertheless, considering the large ellipticity ($\epsilon=0.48$) of Eri~II, we also tested the possibility of a velocity gradient, which could result either from rotational support or a tidal interaction, using a 4-parameter model (i.e., mean velocity $v_{\rm hel}$, velocity dispersion $\sigma_{v_{\rm hel}}$, velocity gradient $\frac{dv}{d\chi}$, and position angle of the gradient $\theta$) similar to that of~\citet{Martin2010} and~\citet{Collins2016}: 

\begin{equation}\label{mle_vdisp}
\medmuskip=-1.5mu
\thinmuskip=-1mu
\thickmuskip=-1mu
\log \mathcal{L} = -\frac{1}{2} \left [  \sum_{n = 1}^{N}\log (\sigma_{v_{\rm hel}}^2 + \sigma_{v_i}^2) +  \sum_{n = 1}^{N} \frac{(v_i - v_{\rm hel} - \frac{dv}{d\chi}\chi_i)^2}{\sigma_{v_i}^2 + \sigma_{v_{\rm hel}}^2} \right ],
\end{equation}\label{eq:mcmc_gradient}

\noindent
where $\chi_i$ is the angular distance between the Eri~II center ($\alpha_0, \delta_0$) and $i$-th star ($\alpha_i, \delta_i$) projected to the gradient axis at a position angle $\theta$:

\begin{equation}\label{mle_vdisp}
\medmuskip=-1mu
\chi_i = (\alpha_i - \alpha_0)\cos(\delta_0)\sin(\theta) + (\delta_i - \delta_0)\cos(\theta).
\end{equation}\label{eq:mcmc_chi}

\noindent
Note that we did not include the astrometric uncertainties of the stars in the likelihood, as the astrometric uncertainties are negligible compared to the velocity uncertainties (i.e., $\frac{dv}{d\chi} \sigma_{\chi_i} \ll \sigma_{v_i}$). 

\begin{figure*}[th!]
\epsscale{1}
\plotone{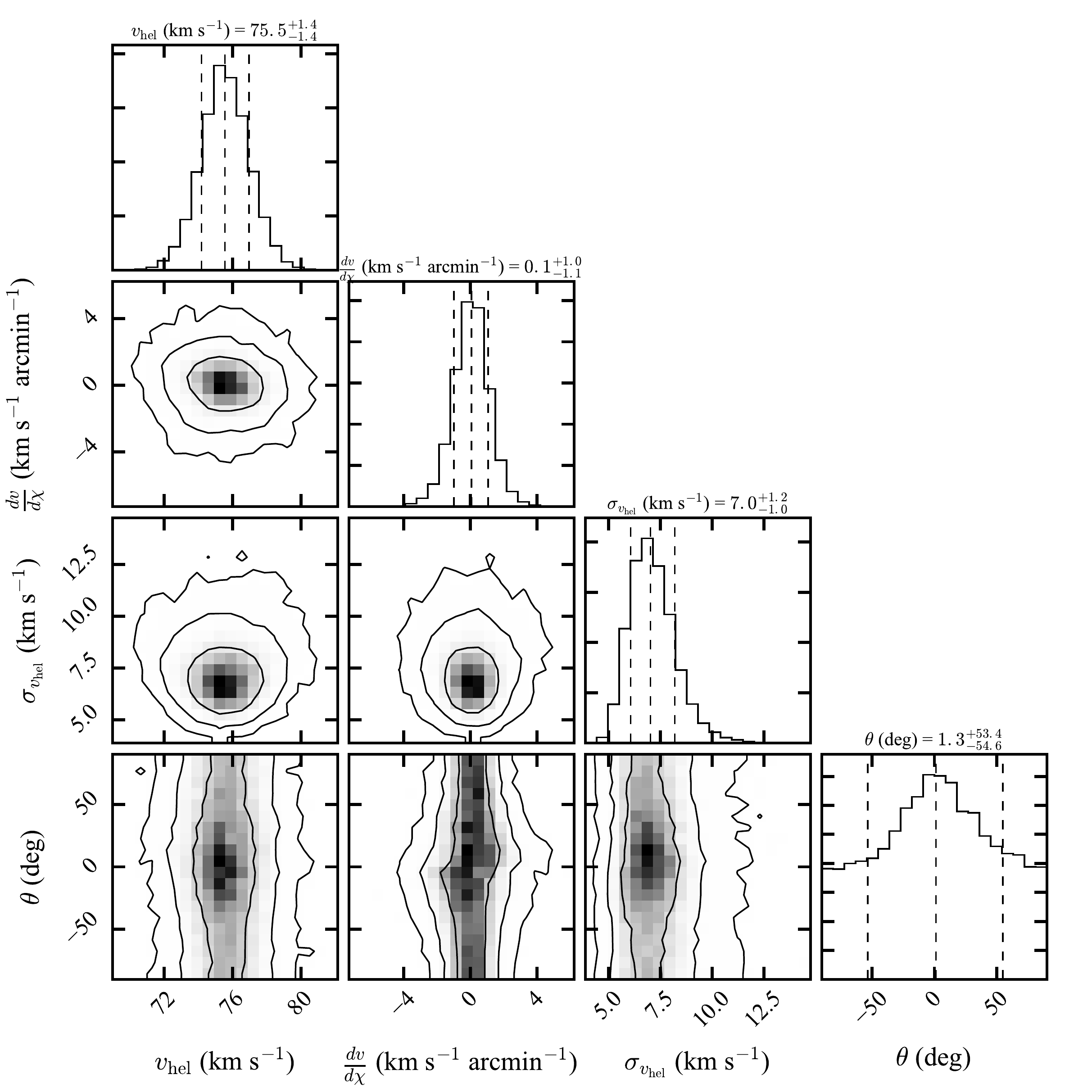}
\caption{Two-dimensional and marginalized posterior probability distribution from a MCMC sampler using a 4-parameter likelihood model defined in the text. The four parameters are mean velocity $v_{\rm hel}$[\kms], velocity dispersion $\sigma_{v_{\rm hel}}$[\kms], velocity gradient $\frac{dv}{d\chi}$ [\kms arcmin$^{-1}$], and position angle of the gradient $\theta$[deg], respectively. Dashed lines and contours have the same meaning as in Figure~\ref{vmcmc}. We conclude from this analysis that there is no evidence for a velocity gradient in Eri~II.}
\label{mcmc_gradient}
\end{figure*}

We ran a 4-parameter MCMC sampler using Equations (3) and (4) to obtain the posterior probability distribution displayed in Figure~\ref{mcmc_gradient}.
Since the best-fit velocity gradient, $\frac{dv}{d\chi} = 0.1 \pm 1.1$~\kms~arcmin$^{-1}$, is consistent with zero within 1$\sigma$ uncertainty, we conclude that there is no evidence for rotation or tidal interaction in Eri~II, which validates the assumption used for mass derivation that Eri~II is a dispersion-supported system in dynamical equilibrium.

\subsection{Metallicity Dispersion}

Using the CaT metallicity measurements, we find that the 16 brightest member stars of Eri~II span more than 1 dex in iron abundance, ranging from $\feh = -1.7$ to $\feh= -3.4$.
We used a similar Gaussian likelihood model to the one described in \S\ref{sec:kinematics} to calculate the mean metallicity and metallicity dispersion of Eri~II.  We find a mean metallicity of $\feh = \fehmedian$, with a dispersion of $\sigma_{\feh} = \fehdisp$.  The posterior probability distributions from the MCMC sampler are displayed on the right side of Figure~\ref{vmcmc}. The kinematic and chemical properties of Eri~II are summarized in Table~\ref{eri2_table}.


\subsection{The Classification of Eridanus~II}\label{nature}

The mass-to-light ratio we have derived for Eri~II indicates that it is a dark matter-dominated dwarf galaxy. This value is consistent with the relation between mass-to-light ratio and luminosity for other dwarf galaxies in the Milky Way and the Local Group \citep[e.g.,][]{sg07}. The low average metallicity (\fehmedian) and large metallicity dispersion (\fehdisp) matches with observations of other dwarf galaxies with similar luminosities~\citep{kirby13b}. 
Combining these results with the orbit discussed in~\S\ref{orbit} and the distance, we conclude that Eri~II is one of the Milky Way's most distant satellite galaxies detected.

\subsection{The Orbit of Eridanus~II}\label{orbit}

The heliocentric velocity $v_{\rm hel} = \vbulknoerr$~\kms corresponds to a velocity in the Galactic Standard of Rest frame\footnote{We adopted the circular orbital velocity of Milky Way at the Sun's radius $\Theta_{0} = 218~\kms$~\citep{Bovy2012} and solar motion of $(U_\odot,~V_\odot,~W_\odot) = (11.1,~12.24,~7.25)~\kms$~\citep{Schonrich2010} for the velocity transformation from heliocentric to Galactic Standard of Rest.} $v_{\rm GSR} = \vgsr$~\kms.  Eri~II is therefore moving toward us, indicating that it is either on its first infall into the Milky Way potential, or it has passed the apocenter of the orbit on a subsequent passage.
To assess whether or not Eri~II is bound to the Milky Way, we derive the escape velocity at the location of Eri~II ($\sim370~\kpc$) by modeling the dark matter halo of the Milky Way as a Navarro-Frenk-White (NFW) profile \citep{nfw96} with a virial mass of $M_{\rm vir} = 10^{12}\msun$ and a concentration of $c = 12$. 
We find that the escape velocity at the location of Eri~II is $\sim200$~\kms and that Eri~II is very likely bound to the Milky Way.
While it is possible that Eri~II has a very large tangential velocity ($\gtrsim 190$~\kms), we find that this situation is unlikely due to the results of simulations discussed below.

To infer the orbital parameters and infall time of Eri~II, we search for Eri~II analogs in Exploring the Local Volume in Simulations~\citep[ELVIS; ][]{Garrison-Kimmel2014}, a suite of cosmological zoom-in $N$-body simulations using a WMAP7~\citep{Komatsu2011} $\Lambda$CDM cosmology.
ELVIS includes 24 isolated dark-matter halos of masses similar to the Milky Way ($M_{\rm vir} = 1.0 - 2.8 \times 10^{12}\msun$ and $R_{\rm vir} = 260 - 360~\kpc$).\footnote{Note that ELVIS has a total of 48 host halos and the other 24 halos are in pairs that resemble the masses, distance, and relative velocity of the Milky Way - Andromeda pair. We did not use those paired halos in this analysis.}

\begin{figure*}[th!]
\epsscale{1}
\plotone{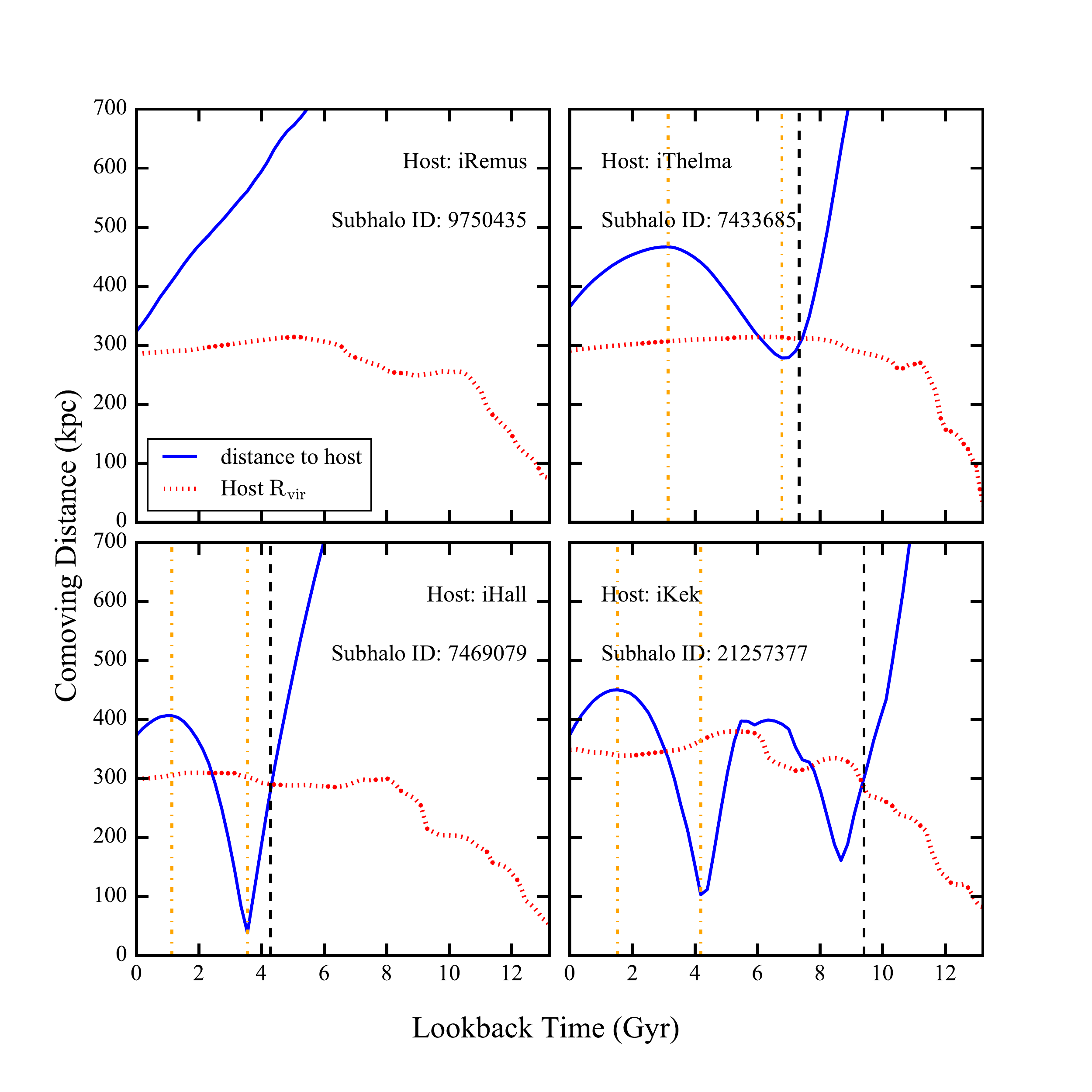}
\caption{Examples of subhalos in the ELVIS simulations that have distances and velocities similar to Eri~II. These subhalos represent four different infall histories for Eri II -- first infall (upper left), second passage with low eccentricity  (upper right), second passage with high eccentricity (lower left), and third passage (lower right). For each panel, the name of the Milky-Way-like host halo and the ID of the subhalo from ELVIS are displayed. The blue solid lines show the (comoving) distance of the subhalo to the host halo and the red dotted lines show the change in the virial radius of the subhalo, both as a function of cosmic time (with time running from right to left). The black dashed lines indicate the infall time of the subhalo and the yellow dashed-dotted lines represent when the subhalo reached its pericenter and apocenter.  About half of the Eri~II analogs found in ELVIS simulation are on their second passage with high eccentric orbit (i.e., similar to the lower left panel); we conclude that this is the most likely scenario for Eri~II infall history. However, proper motion measurements are necessary to better constrain the orbit of Eri~II.
}
\label{elvis}
\end{figure*}

We selected Eri~II analogs that are similar to Eri~II in both the Galactocentric distance and line-of-sight velocity at redshift $z=0$ and are not located inside of any other halos. In order to calculate the line-of-sight velocity, we approximate the observer as being at the center of the host halo. This is a fair approximation as the Galactocentric distance for Eri~II is much larger than Sun's distance to the Milky Way center. Among the 24 ELVIS halos, we find 58 subhalos that possess a similar Galactocentric distance (between 320~\kpc and 420~\kpc) and a similar $v_{\rm GSR}$ (between $-85$~\kms and $-50$~\kms) to Eri~II and have a stellar mass in the range of $M_* = 6\times 10^3-10^5~\msun$.\footnote{Stellar masses in the ELVIS simulations are derived from an abundance matching relation and might have large uncertainties~\citep{Garrison-Kimmel2014}. We therefore accepted a large stellar mass range for the Eri~II analogs.} All 58 of these Eri~II analog subhalos have a binding energy larger than the kinetic energy and therefore they are all bound to their host halos. We therefore conclude that Eri~II is very likely bound to the Milky Way. Among these 58 subhalos, 9 of them are on their first infall (i.e., the subhalos have never entered the virial radius of the host), 41 of them are on their second passages (i.e., the subhalos just passed the apocenter for the first time) with infall time around $4-7$~Gyr ago, and 8 of them are on their third passages (i.e., just passed the apocenter for the second time) with infall time $8-10$~Gyr ago. The second passage cases can also be divided into two categories: 12 subhalos have low eccentricity orbits with pericenter $> 200~\kpc$ and orbital period of $\sim6-7$~Gyr, while 29 subhalos have high eccentricity orbits with pericenter $< 200~\kpc$ and orbital period of $\sim4-6$~Gyr. We therefore conclude that Eri~II is most likely on its second passage, with an eccentric orbit and a first infall time of $\sim5$~ Gyr ago (with a probability of $\sim50\%$). Examples of the infall history in each category are given in Figure~\ref{elvis}.

\subsection{Star Formation in Eridanus~II}\label{quenching}

\citet{koposov15} noted the presence of seven bright ($g \sim 20.5$) blue stars near the center of Eri~II, with colors and magnitudes consistent with being blue loop stars from a $\sim$250~Myr old population in Eri~II (see the stars around $g-r\sim0.3$ and $g\sim20.5$ in left panel of Figure ~\ref{cmd}).  We obtained spectra of five of these stars, and our measurements show that none of them have a velocity close to the systemic velocity of Eri~II. We therefore conclude that the location of these stars near Eri~II is a coincidence, and that there is no evidence for recent star formation.  This result is consistent with the star formation history derived by \citet{Crnojevic16} as well as the low \ion{H}{1} gas content measured by \citet{Westmeier2015} and \citet{Crnojevic16}. In addition to the possible young population,  \citet{Crnojevic16} also identified a possible intermediate-age (3~Gyr) population in Eri~II. However, our spectroscopic measurements are not deep enough to target any of those stars. 

\begin{figure}[th!]
\epsscale{1.24}
\plotone{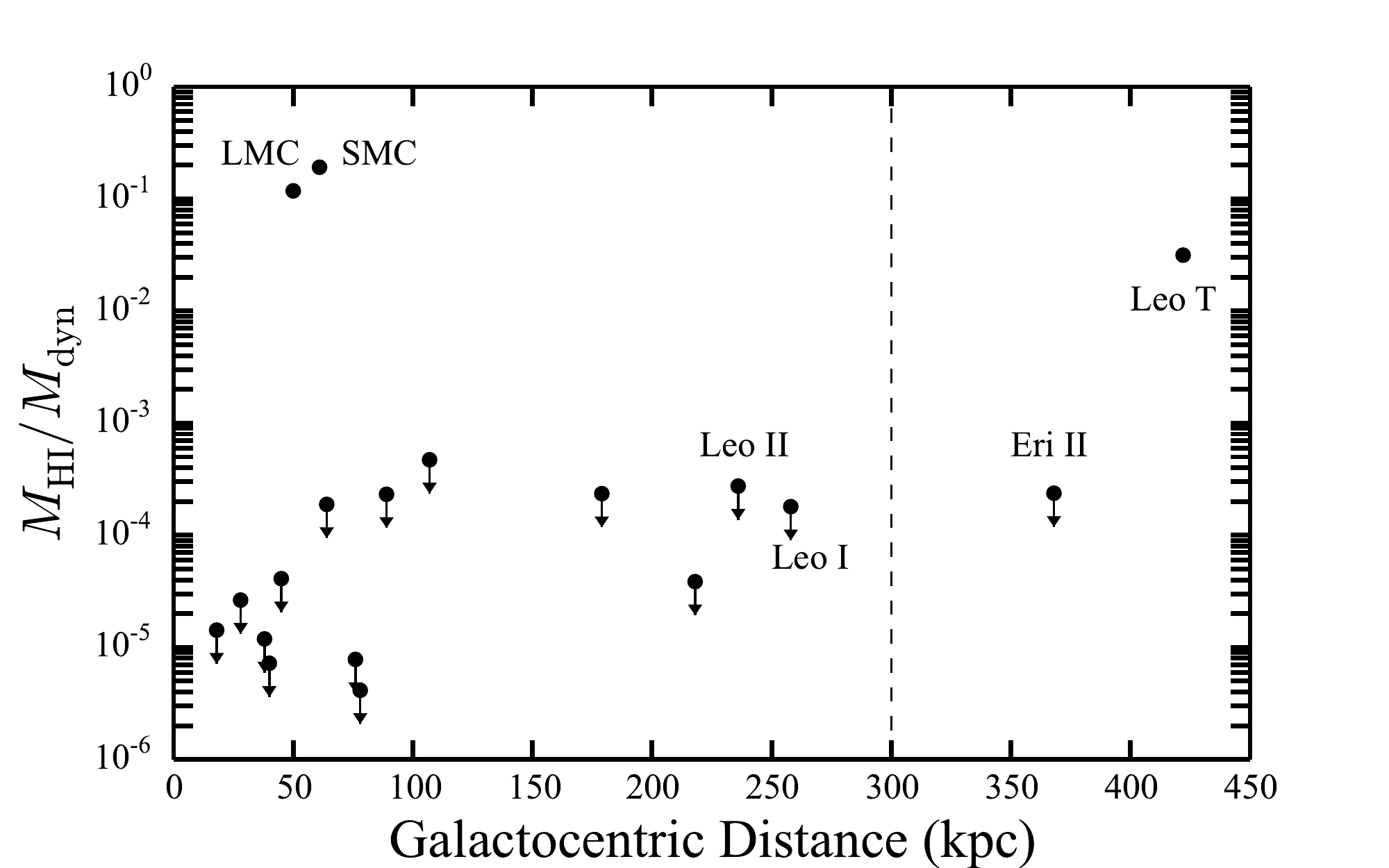} 
\caption{
Ratio of \ion{H}{1} mass, $M_{\rm HI}$, to dynamical mass, $M_{\rm dyn}$, for Milky Way satellites as a function of the Galactocentric distance. The LMC, SMC, and Leo~T are the only Milky Way satellites that have detected neutral gas. The upper limits on \ion{H}{1} mass are used for other satellites. The \ion{H}{1} masses (or upper limits) are taken from~\citet{Crnojevic16} for Eri~II, from~\citet{Ryan-Weber2008} for Leo~T, from~\citet{Bruns2005} for the LMC and SMC, and from~\citet{Spekkens2014} for the remaining dwarfs. For the LMC and SMC, the dynamical mass is adopted from~\citet{vdm02} and~\citet{Stanimirovic2004}. For the pressure-supported systems, the dynamical mass is computed as the mass enclosed within the half-light radius by adopting the formula from~\citet{wolf10}, using the velocity dispersion and half-light radius from~\citet{McConnachie12} (except for Eri~II, which is from this paper). The vertical dashed line shows the approximate virial radius of the Milky Way, $R_{\rm vir} \sim 300$~kpc.
}
\label{h1mass}
\end{figure}

As noted by many previous studies, the dwarf galaxies around the virial radius of the Milky Way show a sharp transition in star formation rate and cold gas content, with the Magellanic Clouds as the only gas-rich, star-forming galaxies inside the Milky Way's virial radius \citep[e.g.,][]{Einasto1974,Blitz2000,Grcevich2009,Spekkens2014}.  Until now, Leo~T ($d\sim420$~kpc) has been the closest known low-luminosity dwarf outside the virial radius of the Milky Way.  Notably, Leo~T retains a significant \ion{H}{1} gas reservoir ~\citep{Irwin2007,Ryan-Weber2008} and has signatures of star formation within the past few hundred Myr~\citep{deJong2008, Weisz2012}, but the large gap in distance between the most distant gas-free objects (Leo~I and Leo~II at $225-250$~kpc) and Leo~T limits its utility in constraining gas loss mechanisms for Milky Way satellites (see Figure~\ref{h1mass}).
Eri~II is located in this gap, slightly beyond the virial radius, and has a similar luminosity to Leo~T and a larger dynamical mass.  It is therefore striking that Eri~II has a much lower gas content and apparently lacks any recent star formation.
As suggested by~\citet{Wetzel2015}, quenching at stellar mass $M_*=10^{4-5}\msun$ may arise from a mix of the host-halo environment and cosmic reionization. 
As a quiescent dwarf galaxy located close to, but beyond, the virial radius of Milky Way, Eri~II is a key object for studying environmental influences on low-mass galaxies and the quenching of star formation in such systems~\citep{weisz14a, Wetzel2015, Wheeler2015, Fillingham2016}.

We compare Eri~II with Leo~T by searching for Leo~T analogs in the ELVIS simulations, as we did for Eri~II in \S\ref{orbit}. The radial velocity of Leo~T~\citep[$v_{\rm GSR} = -58$~\kms;][]{sg07} is similar to that of Eri~II. However, as Leo~T is about 50~kpc farther from the Milky Way than Eri~II, we found that $\sim$30\% of Leo~T analogs are on their first infall into the Milky Way, compared to only $\sim$15\% of Eri II analogs.
If the quenching of Eri~II was caused by host-halo interactions, the different orbits and infall histories of Eri~II and Leo~T could be responsible for their differing star formation histories.
If Eri~II is indeed on its second orbit around the Galaxy as suggested from the results of ELVIS simulation, then the gas reservoir of Eri~II could have been swept away via ram pressure stripping during its first passage. The quenching timescale for galaxies with stellar mass $M_* < 10^7\msun$ is less than 1.5~Gyr~\citep{Fillingham2015, Wetzel2015}, matching well with the possible intermediate-age (3~Gyr) population found by \citet{Crnojevic16}. However, complete removal of the neutral gas during a single orbit around the Milky Way that likely does not closely approach the Galactic disk would place stringent constraints on the halo gas density at large Galactocentric radii. Future proper motion measurements of Eri~II and Leo~T from {\em HST} will better constrain their orbits, determining whether or not they are on their first infall and determining their orbital pericenters.


Alternatively, the quenching of Eri~II could have been caused by cosmic reionization at high redshift~\citep[e.g.,][]{Bullock2001, brown14}. Reionization can explain the lack of gas in Eri~II even if it is on its first infall. However, in that case, Eri~II should not show an extended star formation history or an intermediate-age population. Future deep imaging data from {\em HST} will help determine whether star formation in Eri~II ended very early, as in other ultra-faint dwarfs, or continued to later times.  We note that Eri~II is more distant and more massive than any of the Milky Way satellites strongly suspected to be quenched by reionization based on previous deep {\em HST} imaging \citep{brown14}.
In this scenario, the key question is why Eri~II was more susceptible to the effects of reionization than Leo~T, which is currently less massive than Eri~II but still contains gas.  One possible explanation for the contrast between the observed properties of the two systems is that Leo~T was farther away from the proto-Milky Way at the time of reionization. This hypothesis can be tested with proper motion measurements. If Leo~T first fell into the Milky Way much later than Eri~II, it is also possible that its isolation allowed a late phase of gas accretion and associated star formation~\citep{Ricotti2009}.

\subsection{Constraints on the Nature of Dark Matter}

Ultra-faint dwarf galaxies are ideal targets for understanding the nature of dark matter. They can provide strong tests of models where dark matter is composed of weakly interacting massive particles (WIMPs) that self-annihilate to produce gamma rays~\citep{gunn78,bergstrom88,Baltz:2008wd}. The predicted gamma-ray signal from annihilation is proportional to the line-of-sight integral through the square of the dark matter density, or so-called \Jfactor. The \Jfactor is derived by modeling the velocity using the spherical Jeans equation with prior assumptions on the parameterization of the dark matter halo profile.\citep[\eg,][]{Strigari:2007at,Essig:2009jx,Charbonnier:2011ft,Martinez2015,Geringer-Sameth:2014yza}. Following the procedure of~\citet{simon15b}, we model the dark matter halo as a generalized NFW profile \citep{nfw96}. We use flat, `uninformative' priors on the dark matter halo parameters \citep[see, \eg,][]{Essig:2009jx} and assume a constant stellar velocity anisotropy. We find an integrated \Jfactor for Eri~II of $\log_{10}(J) = \jsmall$ within solid angle of $0.2\degr$, and $\log_{10}(J) = \jlarge$ within $0.5\degr$. The error bars represent the difference between the 16th and 84th percentiles and the median of the posterior distribution of \Jfactor. These values assume that the dark matter halo extends beyond the radius of the outermost spectroscopically confirmed star, but truncates within the estimated tidal radius for the dark matter halo. Given that Eri II is $\sim 370~\kpc$ away, the tidal radius could extend far beyond its outermost star location. We derive a tidal radius of $\sim10~\kpc$ following the description in \citet{Geringer-Sameth:2014yza}. The \Jfactor of Eri~II is $\sim 3$ orders of magnitude smaller than the most promising dwarf galaxies, which is a direct result of the distance between Eri~II and the Sun. This value is consistent with the value predicted from a simple distance scaling based on the \Jfactors of known dwarfs~\citep[e.g.,][]{dw15a}.

While Eri~II does not appear to be a promising target for indirect searches for WIMP annihilation due to its distance, the existence of a central star cluster offers a unique opportunity to constrain massive compact halo object (MACHO) dark matter with mass $\gtrsim 10$~\msun \citep{Brandt2016}.
The recent detection of gravitational waves from the coalescence of $\sim$30~\msun black holes \citep{LIGO2016} has led to the suggestion that primordial black holes with a similar mass could constitute the dark matter \citep[e.g.,][]{Bird2016,Clesse2016}.
Interestingly, microlensing and wide binaries searches do not exclude MACHOs in the mass range between 20~\msun and 100~\msun~\citep[e.g.,][]{Alcock2001,Tisserand2007,Quinn2009}. 
We therefore examine the constraints that can be placed on MACHO dark matter using the measured properties of Eri~II.

\begin{figure}[th!]
\epsscale{1.24}
\plotone{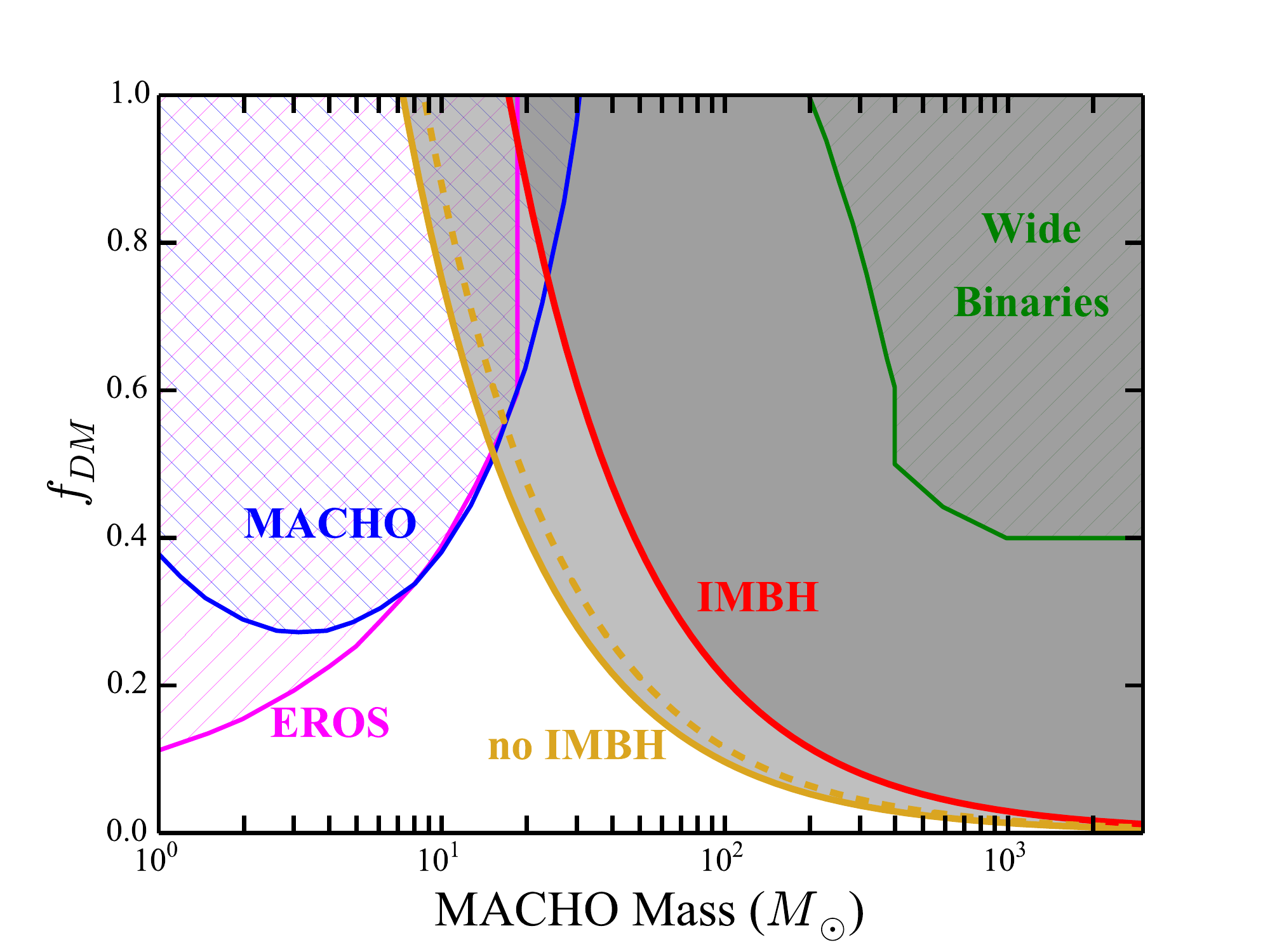}
\caption{
Constraints on MACHO dark matter following the prescription of \citet{Brandt2016}, assuming that the Eri~II star cluster is located at the center of the Eri~II dark matter potential. Colored curves mark exclusion regions for the maximum fraction of dark matter ($f_{\rm DM}$) in MACHOs for a given MACHO mass. The solid yellow curve corresponds to the limits derived from the observed 3D velocity dispersion of $\sigma_{3D} = 12~\kms$ and implied central dark matter density of $\rho = 0.15~M_\odot \pc^{-3}$. As a comparison, the limit derived from $\sigma_{3D} = 8~\kms$ and  $\rho = 0.02~M_\odot \pc^{-3}$ from \citet{Brandt2016} is shown as the dashed yellow curve. Since the increase in $\sigma_{3D}$ makes the constraint looser (i.e., shifts the curve rightward) and the increase in $\rho$ makes the constraint stronger (i.e., shifts the curve leftward), the combination of the two leads to similar results despite different $\sigma_{3D}$ and $\rho$ values. The red curve shows the MACHO constraint assuming that an intermediate mass black hole (IMBH) with mass of $1500~M_\odot$ resides at the center of Eri~II. Magenta and blue hatched contours correspond to microlensing constraints from the EROS~\citep{Tisserand2007} and MACHO~\citep{Alcock2001} experiments, respectively. The green hatched curve corresponds to constraints from the stability of wide binary stars~\citep{Quinn2009}. Note that the MACHO limits from these experiments extend to much smaller and larger masses than are displayed in this plot.
}
\label{macho}
\end{figure}

\citet{Brandt2016} argued that MACHO dark matter would dynamically heat, and eventually dissolve, the star cluster near the center of Eri~II. Brandt projected MACHO constraints from the survival of this star cluster assuming several values for the three-dimensional velocity dispersion, $\sigma_{\rm 3D}$, and dark matter density, $\rho$, of Eri~II.  However, the kinematics measured in \S\ref{sec:kinematics} allowed us to directly derive the three-dimensional velocity dispersion and dark matter density of Eri~II: $\sigma_{\rm 3D} \sim 12$~\kms
and $\rho \sim 0.15~\msun/\pc^3$,\footnote{Here we assume that velocity dispersion for the dark matter halo is the same as the velocity dispersion for the stars, and $\sigma_{\rm 3D} = \sqrt{3}\sigma_{\rm 1D}$.} assuming a uniform and isotropic distribution of dark matter within the half-light radius. 
With these halo properties, we derived MACHO constraints assuming the stellar cluster has an age of 3 Gyr, an initial half-light radius of $r_{h,0}\sim13$~pc, and a mass of 2000~\msun,\footnote{The star cluster has an absolute magnitude of $M_v = -3.5$ ($\sim$2000~\lsun) and half-light radius $r_{h,\mathrm{cluster}} = 13$~pc~\citep{Crnojevic16}. The assumptions for the stellar cluster are based on these observational results and the 3~Gyr intermediate-age population found in Eri~II~\citep{Crnojevic16}. We note that an older population for the cluster is possible, which would lead to a stronger MACHO constraint (i.e., shift the curve leftward).} as shown in Figure~\ref{macho}. 
When the results from Eri~II are combined with those from microlensing and wide binary searches~\citep{Alcock2001,Tisserand2007,Quinn2009}, MACHOs with mass $\gtrsim 10^{-7}\msun$ are constrained to be a subdominant component of dark matter.
However, if there were an intermediate mass black hole (IMBH) of mass $M_{\rm BH}=1500~M_\odot$ at the center of Eri~II, as extrapolation of the scaling relation from~\citet{Kruijssen2013} suggests, its gravity would stabilize the star cluster and would prevent its evaporation. This effect would weaken the bounds and allow for a MACHO mass distribution peaked at a few tens of solar masses to be the main component of the dark matter in the universe~\citep{Clesse2015}.

In addition to constraints on MACHOs, Eri~II may also offer an opportunity to constrain the density profile of dark matter halos, addressing the so-called ``cusp-core problem''. 
Historically, cosmological simulations~\citep{Dubinski1991,nfw96} predict that dark matter halos should have a cuspy central density profile.  However, observational results have consistently pointed to shallower profiles~\citep[\eg,][]{deblok01,walker:2011zu,adams14}, and theoretical calculations have been developed to explain why cusps are not found~\citep[\eg,][]{governato12,pontzen12,DiCintio2014}.  In these models, dark matter cores arise from gravitational interactions between dark matter and baryons, and should occur in dwarf galaxies that underwent multiple vigorous episodes of star formation. \citet{Crnojevic16} show that Eri~II is possibly the least massive dwarf galaxy that is known to have an extended star formation history and therefore its density profile may also be affected by baryons. The star cluster of Eri~II may offer potential to constrain the dark matter profile of Eri~II through survivability arguments~\citep[see, e.g.,][]{Cole2012} and could provide an independent probe of the dark matter profile shape. A better understanding of the dark matter distribution at small scales will help us understand how the dwarf galaxies we observe today are linked to the primordial population of dark matter subhalos predicted by $\Lambda$CDM cosmology.

\section{SUMMARY}
\label{conclusions}

We obtained Magellan/IMACS spectroscopy of stars in the recently-discovered Milky Way satellite Eridanus~II. We identified 28 members based on the radial velocities of 54 stars in the vicinity of Eri~II. Of the confirmed members, 16 stars have large enough S/N to measure their metallicity. 

We find a systemic velocity of $v_{\rm hel} = \vbulk$~\kms\ ($v_{\rm GSR} = \vgsr$~\kms) and a velocity dispersion of \vdisp~\kms.  The mass within  the half-light radius of Eri~II is $M_{\rm 1/2} = \mass ~\msun$, corresponding to a dynamical mass-to-light ratio of 
\masstolight~\msun/\lsun. The mean metallicity of Eri~II is $\feh = \fehmedian$, with a metallicity dispersion of $\sigma_{\feh} = \fehdisp$. Both the dynamical and chemical evidence show that Eri~II is a dark matter-dominated dwarf galaxy. 

The negative velocity of Eri~II in the Galactic Standard of Rest frame implies that it is moving towards the Milky Way. Eri~II is therefore either falling into the Milky Way potential for the first time, or it has recently passed the apocenter of its orbit. By identifying subhalos in dark matter-only simulations that are consistent with the line-of-sight velocity and distance of Eri~II, we show that Eri~II is very likely bound to the Milky Way and it is mostly likely on its second infall with an eccentric orbit. Future measurements of its proper motion will better constrain its orbit and confirm its origin. Furthermore, our measurements of radial velocities show that none of the candidate blue loop stars we observed are associated with Eri~II. We therefore conclude that there is no evidence for recent star formation ($\sim$250~Myr) in Eri~II. 

Though Eri~II is not a promising target for indirect searches for WIMP annihilation due to its distance, it offers a unique opportunity to constrain MACHO dark matter because of the existence of a central star cluster. Our spectroscopic analysis provides a direct measurement of the mass and density of the Eri~II's dark matter halo, and therefore provides more precise constraints on the abundance of MACHO dark matter. Moreover, the existence of the star cluster may also offer an independent probe to constrain the dark matter density profile.

\acknowledgements{
This paper has gone through internal review by the DES collaboration. We thank the anonymous referee for suggestions that substantially improved the quality of the paper.
T.S.L. and J.D.S. thank Dan Kelson for many helpful conversations regarding IMACS data reduction.
T.S.L. acknowledges support by the Mitchell Institute Fellowship from Texas A\&M University and the Leon Lederman Fellowship from Fermilab. 
J.D.S. acknowledges support from the National Science Foundation under grant AST-1108811. 
J.G.B. acknowledges support from Centro de Excelencia Severo Ochoa SEV-2012-0234 and SEV-2012-0249.
E.B. acknowledges financial support from the European Research Council (ERC-StG-335936).

This research has made use of NASA's Astrophysics Data System Bibliographic Services.
This research made use of \code{Astropy}, a community-developed core Python package for Astronomy~\citep{Astropy2013}.
Contour plots were generated using \code{corner.py} \citep{corner}.

Funding for the DES Projects has been provided by the U.S. Department of Energy, the U.S. National Science Foundation, the Ministry of Science and Education of Spain, 
the Science and Technology Facilities Council of the United Kingdom, the Higher Education Funding Council for England, the National Center for Supercomputing 
Applications at the University of Illinois at Urbana-Champaign, the Kavli Institute of Cosmological Physics at the University of Chicago, 
the Center for Cosmology and Astro-Particle Physics at the Ohio State University,
the Mitchell Institute for Fundamental Physics and Astronomy at Texas A\&M University, Financiadora de Estudos e Projetos, 
Funda{\c c}{\~a}o Carlos Chagas Filho de Amparo {\`a} Pesquisa do Estado do Rio de Janeiro, Conselho Nacional de Desenvolvimento Cient{\'i}fico e Tecnol{\'o}gico and 
the Minist{\'e}rio da Ci{\^e}ncia, Tecnologia e Inova{\c c}{\~a}o, the Deutsche Forschungsgemeinschaft and the Collaborating Institutions in the Dark Energy Survey. 

The Collaborating Institutions are Argonne National Laboratory, the University of California at Santa Cruz, the University of Cambridge, Centro de Investigaciones Energ{\'e}ticas, 
Medioambientales y Tecnol{\'o}gicas-Madrid, the University of Chicago, University College London, the DES-Brazil Consortium, the University of Edinburgh, 
the Eidgen{\"o}ssische Technische Hochschule (ETH) Z{\"u}rich, 
Fermi National Accelerator Laboratory, the University of Illinois at Urbana-Champaign, the Institut de Ci{\`e}ncies de l'Espai (IEEC/CSIC), 
the Institut de F{\'i}sica d'Altes Energies, Lawrence Berkeley National Laboratory, the Ludwig-Maximilians Universit{\"a}t M{\"u}nchen and the associated Excellence Cluster Universe, 
the University of Michigan, the National Optical Astronomy Observatory, the University of Nottingham, The Ohio State University, the University of Pennsylvania, the University of Portsmouth, 
SLAC National Accelerator Laboratory, Stanford University, the University of Sussex, Texas A\&M University, and the OzDES Membership Consortium.

The DES data management system is supported by the National Science Foundation under Grant Number AST-1138766.
The DES participants from Spanish institutions are partially supported by MINECO under grants AYA2012-39559, ESP2013-48274, FPA2013-47986, and Centro de Excelencia Severo Ochoa SEV-2012-0234 and SEV-2012-0249, some of which include ERDF funds from the European Union..
Research leading to these results has received funding from the European Research Council under the European Union’s Seventh Framework Programme (FP7/2007-2013) including ERC grant agreements 240672, 291329, and 306478.
}

{\it Facilities:} 
\facility{This paper includes data gathered with the 6.5 meter
  Magellan Telescopes located at Las Campanas Observatory, Chile.}

\bibliographystyle{apj}
\bibliography{main}{}

\clearpage 
\LongTables

\tabletypesize{\scriptsize}
\begin{deluxetable*}{c c c c c c c r c c c}
\tablecaption{Velocity and metallicity measurements for Tucanan III.\label{tab:eri2_spec}}

\tablehead{ID & MJD\tablenotemark{a} & RA & DEC & $g$\tablenotemark{b} & $r$\tablenotemark{b} & S/N & \multicolumn{1}{c}{$v$} & ${\rm EW}$ & ${\rm [Fe/H]}$ & MEM \\
 &  & (deg) & (deg) & (mag) & (mag) &  & \multicolumn{1}{c}{(\kms)} & (\AA) &  &  }
\startdata

DES\,J034338.10$-$432550.9 & 57345.7 & 55.90874 & -43.43079 & 22.23 & 20.82 & 15.4 & $17.39 \pm 1.14$ & $4.61 \pm 0.50$ & -- &    0\\
... & 57312.8 & ... & ... & ... & ... &  7.0 & $14.50 \pm 1.99$ & $3.07 \pm 1.80$ & -- & ...\\
DES\,J034340.06$-$432808.3 & 57345.7 & 55.91690 & -43.46898 & 20.14 & 18.82 & 59.9 & $20.12 \pm 1.01$ & $5.85 \pm 0.24$ & -- &    0\\
... & 57312.8 & ... & ... & ... & ... & 29.2 & $19.64 \pm 1.24$ & $5.90 \pm 0.34$ & -- & ...\\
DES\,J034341.56$-$432918.2 & 57345.7 & 55.92317 & -43.48838 & 19.97 & 18.45 & 139.8 & $52.41 \pm 1.00$ & $2.15 \pm 0.20$ & -- &    0\\
... & 57312.8 & ... & ... & ... & ... & 70.2 & $54.07 \pm 1.21$ & $2.30 \pm 0.21$ & -- & ...\\
DES\,J034343.30$-$432810.1 & 57345.7 & 55.93042 & -43.46948 & 20.63 & 19.22 & 68.4 & $13.19 \pm 1.01$ & $3.55 \pm 0.23$ & -- &    0\\
... & 57312.8 & ... & ... & ... & ... & 32.4 & $13.89 \pm 1.23$ & $3.50 \pm 0.32$ & -- & ...\\
DES\,J034347.77$-$432951.7 & 57345.7 & 55.94904 & -43.49771 & 20.53 & 20.21 & 13.2 & $306.12 \pm 1.27$ & $4.43 \pm 0.45$ & -- &    0\\
... & 57312.8 & ... & ... & ... & ... &  5.7 & $308.51 \pm 2.30$ & -- & -- & ...\\
DES\,J034349.44$-$432432.1 & 57345.7 & 55.95601 & -43.40893 & 20.23 & 18.85 & 69.9 & $105.33 \pm 1.01$ & $5.05 \pm 0.23$ & -- &    0\\
... & 57312.8 & ... & ... & ... & ... & 33.8 & $103.55 \pm 1.26$ & $5.55 \pm 0.31$ & -- & ...\\
DES\,J034349.53$-$432609.3 & 57345.7 & 55.95637 & -43.43590 & 21.58 & 20.48 & 14.8 & $219.50 \pm 1.42$ & -- & -- &    0\\
... & 57312.8 & ... & ... & ... & ... &  7.1 & $220.91 \pm 2.02$ & -- & -- & ...\\
DES\,J034349.54$-$432951.6 & 57345.7 & 55.95642 & -43.49765 & 20.64 & 19.17 & 75.9 & $42.94 \pm 1.01$ & $3.82 \pm 0.22$ & -- &    0\\
... & 57312.8 & ... & ... & ... & ... & 37.9 & $44.60 \pm 1.22$ & $3.68 \pm 0.30$ & -- & ...\\
DES\,J034351.43$-$432639.7 & 57345.7 & 55.96431 & -43.44436 & 21.74 & 20.25 & 26.8 & $-2.63 \pm 1.04$ & $4.54 \pm 0.22$ & -- &    0\\
... & 57312.8 & ... & ... & ... & ... & 11.7 & $-3.70 \pm 1.46$ & $4.76 \pm 0.66$ & -- & ...\\
DES\,J034351.50$-$432714.9 & 57345.7 & 55.96460 & -43.45415 & 19.55 & 18.94 & 37.6 & $46.49 \pm 1.02$ & -- & -- &    0\\
... & 57312.8 & ... & ... & ... & ... & 17.9 & $49.18 \pm 1.31$ & $7.14 \pm 0.51$ & -- & ...\\
DES\,J034355.42$-$432447.2 & 57345.7 & 55.98091 & -43.41310 & 20.36 & 20.12 & 14.3 & $150.44 \pm 1.60$ & $2.66 \pm 0.50$ & -- &    0\\
... & 57312.8 & ... & ... & ... & ... &  7.1 & $147.56 \pm 2.13$ & $2.56 \pm 0.59$ & -- & ...\\
DES\,J034402.04$-$432608.5 & 57345.7 & 56.00848 & -43.43570 & 20.98 & 19.56 & 63.8 & $83.30 \pm 1.01$ & $3.72 \pm 0.23$ & -- &    0\\
... & 57312.8 & ... & ... & ... & ... & 30.3 & $82.90 \pm 1.24$ & $3.78 \pm 0.36$ & -- & ...\\
DES\,J034402.24$-$433158.8 & 57345.7 & 56.00932 & -43.53299 & 21.73 & 21.00 &  7.2 & $69.80 \pm 1.61$ & $4.92 \pm 0.77$ & $-1.85 \pm 0.31$ &    1\\
DES\,J034404.78$-$432727.7 & 57345.7 & 56.01991 & -43.45769 & 21.07 & 20.12 & 20.0 & $64.87 \pm 1.43$ & -- & -- &    0\\
... & 57312.8 & ... & ... & ... & ... &  8.1 & $67.53 \pm 3.08$ & -- & -- & ...\\
DES\,J034406.25$-$432811.1 & 57345.7 & 56.02605 & -43.46976 & 22.34 & 21.07 & 11.4 & $116.91 \pm 1.57$ & -- & -- &    0\\
DES\,J034406.86$-$433105.2 & 57345.7 & 56.02857 & -43.51812 & 22.86 & 21.68 &  5.9 & $262.86 \pm 2.84$ & -- & -- &    0\\
DES\,J034406.94$-$433143.4 & 57345.7 & 56.02892 & -43.52871 & 21.11 & 20.16 & 15.2 & $77.77 \pm 1.23$ & -- & -- &    1\\
... & 57312.8 & ... & ... & ... & ... &  7.2 & $77.96 \pm 1.81$ & -- & -- & ...\\
DES\,J034408.52$-$433046.8 & 57345.7 & 56.03551 & -43.51300 & 20.00 & 18.77 & 61.3 & $18.78 \pm 1.01$ & $5.67 \pm 0.24$ & -- &    0\\
... & 57312.8 & ... & ... & ... & ... & 28.9 & $21.47 \pm 1.24$ & $5.23 \pm 0.34$ & -- & ...\\
DES\,J034410.56$-$432602.0 & 57345.7 & 56.04402 & -43.43390 & 20.05 & 18.65 & 97.6 & $39.08 \pm 1.00$ & $4.26 \pm 0.22$ & -- &    0\\
... & 57312.8 & ... & ... & ... & ... & 47.4 & $40.66 \pm 1.21$ & $4.27 \pm 0.26$ & -- & ...\\
DES\,J034411.10$-$433052.1 & 57345.7 & 56.04626 & -43.51447 & 22.02 & 21.28 &  6.5 & $65.37 \pm 2.30$ & -- & -- &    1\\
DES\,J034412.28$-$433105.9 & 57345.7 & 56.05116 & -43.51831 & 21.98 & 21.25 &  5.3 & $75.12 \pm 2.57$ & -- & -- &    1\\
DES\,J034412.37$-$432803.5 & 57345.7 & 56.05156 & -43.46764 & 21.44 & 20.64 & 12.5 & $308.95 \pm 1.20$ & -- & -- &    0\\
DES\,J034412.63$-$433031.3 & 57345.7 & 56.05264 & -43.50870 & 21.79 & 21.10 &  8.2 & $91.01 \pm 1.94$ & -- & -- &    1\\
... & 57312.8 & ... & ... & ... & ... &  3.4 & $91.67 \pm 2.62$ & -- & -- & ...\\
DES\,J034414.62$-$433134.8 & 57345.7 & 56.06092 & -43.52633 & 21.57 & 20.68 & 13.0 & $72.37 \pm 1.54$ & $3.35 \pm 0.49$ & $-2.54 \pm 0.21$ &    1\\
... & 57312.8 & ... & ... & ... & ... &  5.7 & $73.94 \pm 2.08$ & -- & -- & ...\\
DES\,J034415.65$-$433032.0 & 57345.7 & 56.06520 & -43.50890 & 22.09 & 21.31 &  7.4 & $65.75 \pm 1.60$ & -- & -- &    1\\
DES\,J034416.14$-$433243.4 & 57345.7 & 56.06724 & -43.54538 & 20.58 & 19.58 & 28.8 & $74.11 \pm 1.05$ & $3.96 \pm 0.26$ & $-2.54 \pm 0.12$ &    1\\
... & 57312.8 & ... & ... & ... & ... & 14.2 & $73.94 \pm 1.36$ & $4.25 \pm 0.51$ & $-2.43 \pm 0.20$ & ...\\
DES\,J034416.29$-$433038.7 & 57345.7 & 56.06786 & -43.51076 & 20.49 & 20.15 & 13.4 & $199.69 \pm 1.46$ & $2.96 \pm 0.60$ & -- &    0\\
DES\,J034418.18$-$433111.9 & 57345.7 & 56.07574 & -43.51998 & 21.40 & 20.69 & 10.8 & $80.37 \pm 1.70$ & $3.19 \pm 0.50$ & $-2.63 \pm 0.22$ &    1\\
... & 57312.8 & ... & ... & ... & ... &  4.3 & $81.44 \pm 3.16$ & -- & -- & ...\\
DES\,J034419.20$-$433018.9 & 57345.7 & 56.08000 & -43.50525 & 22.36 & 21.68 &  5.4 & $81.69 \pm 2.95$ & -- & -- &    1\\
DES\,J034420.20$-$433210.9 & 57345.7 & 56.08417 & -43.53636 & 21.42 & 20.53 & 12.0 & $76.71 \pm 1.62$ & -- & -- &    1\\
... & 57312.8 & ... & ... & ... & ... &  6.3 & $78.55 \pm 2.14$ & -- & -- & ...\\
DES\,J034420.62$-$433308.1 & 57345.7 & 56.08593 & -43.55225 & 21.71 & 21.02 &  7.4 & $84.54 \pm 2.04$ & -- & -- &    1\\
... & 57312.8 & ... & ... & ... & ... &  3.9 & $79.33 \pm 3.24$ & -- & -- & ...\\
DES\,J034420.77$-$433227.6 & 57345.7 & 56.08655 & -43.54101 & 21.28 & 20.37 & 16.8 & $86.55 \pm 1.60$ & $5.06 \pm 0.49$ & $-1.94 \pm 0.20$ &    1\\
... & 57312.8 & ... & ... & ... & ... &  7.6 & $85.82 \pm 2.02$ & $5.05 \pm 0.75$ & $-1.94 \pm 0.29$ & ...\\
DES\,J034421.34$-$433020.9 & 57345.7 & 56.08892 & -43.50581 & 21.30 & 20.43 & 15.3 & $75.80 \pm 1.19$ & $5.07 \pm 0.52$ & $-1.92 \pm 0.21$ &    1\\
... & 57312.8 & ... & ... & ... & ... &  6.7 & $83.36 \pm 2.42$ & -- & -- & ...\\
DES\,J034423.06$-$433124.4 & 57345.7 & 56.09608 & -43.52345 & 21.67 & 20.94 &  9.5 & $74.74 \pm 1.41$ & $2.63 \pm 0.60$ & $-2.82 \pm 0.28$ &    1\\
... & 57312.8 & ... & ... & ... & ... &  3.6 & $72.81 \pm 2.42$ & -- & -- & ...\\
DES\,J034423.98$-$433243.6 & 57345.7 & 56.09990 & -43.54543 & 21.05 & 20.09 & 19.9 & $67.60 \pm 1.12$ & $4.60 \pm 0.34$ & $-2.18 \pm 0.14$ &    1\\
... & 57312.8 & ... & ... & ... & ... &  9.7 & $67.79 \pm 1.53$ & $3.64 \pm 0.44$ & $-2.55 \pm 0.18$ & ...\\
DES\,J034426.53$-$433243.9 & 57345.7 & 56.11054 & -43.54552 & 21.29 & 20.58 & 12.3 & $68.82 \pm 1.34$ & $5.44 \pm 0.55$ & $-1.76 \pm 0.22$ &    1\\
... & 57312.8 & ... & ... & ... & ... &  4.7 & $71.50 \pm 2.21$ & -- & -- & ...\\
DES\,J034426.64$-$433122.8 & 57345.7 & 56.11100 & -43.52301 & 21.18 & 20.24 & 19.7 & $77.32 \pm 1.19$ & $5.37 \pm 0.40$ & $-1.85 \pm 0.16$ &    1\\
... & 57312.8 & ... & ... & ... & ... &  8.2 & $79.47 \pm 1.60$ & -- & -- & ...\\
DES\,J034427.50$-$433252.8 & 57345.7 & 56.11457 & -43.54801 & 21.98 & 21.34 &  5.1 & $75.07 \pm 2.40$ & $4.07 \pm 0.51$ & $-2.12 \pm 0.22$ &    1\\
DES\,J034428.27$-$433250.7 & 57345.7 & 56.11778 & -43.54742 & 20.75 & 19.77 & 26.9 & $72.18 \pm 1.05$ & $3.78 \pm 0.31$ & $-2.57 \pm 0.13$ &    1\\
... & 57312.8 & ... & ... & ... & ... & 12.4 & $71.13 \pm 1.42$ & $3.46 \pm 0.50$ & $-2.69 \pm 0.21$ & ...\\
DES\,J034429.32$-$433130.3 & 57345.7 & 56.12217 & -43.52509 & 21.41 & 20.71 & 11.3 & $73.91 \pm 1.27$ & $3.12 \pm 0.26$ & $-2.66 \pm 0.12$ &    1\\
... & 57312.8 & ... & ... & ... & ... &  5.2 & $78.05 \pm 2.59$ & -- & -- & ...\\
DES\,J034429.71$-$433147.9 & 57345.7 & 56.12379 & -43.52998 & 21.17 & 20.32 & 17.1 & $70.79 \pm 1.21$ & $1.65 \pm 0.22$ & $-3.42 \pm 0.15$ &    1\\
... & 57312.8 & ... & ... & ... & ... &  7.9 & $72.21 \pm 1.67$ & -- & -- & ...\\
DES\,J034430.00$-$433305.8 & 57345.7 & 56.12499 & -43.55162 & 20.51 & 20.19 & 14.4 & $224.86 \pm 1.46$ & -- & -- &    0\\
DES\,J034430.24$-$433048.0 & 57345.7 & 56.12601 & -43.51333 & 22.41 & 21.75 &  4.6 & $79.29 \pm 2.40$ & -- & -- &    1\\
DES\,J034431.11$-$433316.0 & 57345.7 & 56.12962 & -43.55444 & 21.70 & 20.93 &  9.0 & $89.73 \pm 1.52$ & $4.49 \pm 0.77$ & $-2.04 \pm 0.31$ &    1\\
... & 57312.8 & ... & ... & ... & ... &  4.5 & $89.41 \pm 3.03$ & -- & -- & ...\\
DES\,J034433.36$-$433319.1 & 57345.7 & 56.13898 & -43.55531 & 21.84 & 21.08 &  7.0 & $66.71 \pm 1.98$ & -- & -- &    1\\
DES\,J034435.17$-$433306.2 & 57345.7 & 56.14654 & -43.55172 & 22.39 & 21.00 & 16.2 & $72.36 \pm 1.37$ & $4.54 \pm 0.47$ & -- &    0\\
... & 57312.8 & ... & ... & ... & ... &  7.2 & $72.16 \pm 1.69$ & $4.06 \pm 0.60$ & -- & ...\\
DES\,J034437.06$-$433420.7 & 57345.7 & 56.15440 & -43.57242 & 22.69 & 21.31 & 14.6 & $14.30 \pm 1.16$ & $4.51 \pm 0.56$ & -- &    0\\
... & 57312.8 & ... & ... & ... & ... &  7.2 & $11.95 \pm 1.65$ & -- & -- & ...\\
DES\,J034437.87$-$433457.2 & 57345.7 & 56.15777 & -43.58255 & 21.82 & 21.04 &  5.6 & $362.64 \pm 2.80$ & -- & -- &    0\\
DES\,J034438.15$-$433549.0 & 57345.7 & 56.15897 & -43.59693 & 20.17 & 18.75 & 71.7 & $-1.20 \pm 1.01$ & $4.38 \pm 0.22$ & -- &    0\\
... & 57312.8 & ... & ... & ... & ... & 37.9 & $-1.43 \pm 1.22$ & $4.59 \pm 0.28$ & -- & ...\\
DES\,J034438.78$-$433015.2 & 57345.7 & 56.16156 & -43.50421 & 21.79 & 21.06 &  8.5 & $73.76 \pm 2.09$ & -- & -- &    1\\
... & 57312.8 & ... & ... & ... & ... &  2.6 & $75.78 \pm 3.89$ & -- & -- & ...\\
DES\,J034439.68$-$433038.6 & 57345.7 & 56.16534 & -43.51073 & 21.94 & 21.20 &  5.9 & $68.39 \pm 2.17$ & -- & -- &    1\\
DES\,J034440.26$-$433419.3 & 57345.7 & 56.16773 & -43.57202 & 21.26 & 19.89 & 34.8 & $49.29 \pm 1.04$ & $5.00 \pm 0.29$ & -- &    0\\
... & 57312.8 & ... & ... & ... & ... & 17.9 & $52.49 \pm 1.32$ & $5.05 \pm 0.54$ & -- & ...\\
DES\,J034440.32$-$433336.5 & 57345.7 & 56.16801 & -43.56015 & 20.46 & 20.26 & 11.4 & $-120.75 \pm 1.41$ & -- & -- &    0\\
DES\,J034445.57$-$432955.4 & 57345.7 & 56.18989 & -43.49872 & 21.23 & 20.43 & 16.2 & $82.22 \pm 1.24$ & $2.20 \pm 0.39$ & $-3.12 \pm 0.20$ &    1\\
... & 57312.8 & ... & ... & ... & ... &  7.1 & $76.41 \pm 1.87$ & -- & -- & ...\\

\enddata
\tablenotetext{a}{MJD = 57345.7 corresponds to the November run and MJD = 57312.8 corresponds to the October run. Since for both runs the observations were made over two nights, the MJD listed here is the weighted mean observation date, which occurs during daylight hours.}
\tablenotetext{b}{Quoted magnitudes represent the weighted-average dereddened PSF magnitude derived from the DES images with SExtractor.}
\end{deluxetable*}

\end{document}